\documentclass[aps,prl,reprint,superscriptaddress]{revtex4-2}

\usepackage{graphicx}
\usepackage{dcolumn}
\usepackage{bm,braket,here,comment}
\usepackage{mathtools,amssymb,color,xcolor}
\usepackage[normalem]{ulem}

\allowdisplaybreaks[1]

\renewcommand\>{\rangle}

\begin{document}

\preprint{}

\title{Observation of slow relaxation due to Hilbert space fragmentation\\ in strongly interacting Bose-Hubbard chains}


\author{Kantaro Honda}
 \email[]{honda.kantaro.35m@st.kyoto-u.ac.jp}
\altaffiliation{}
 \affiliation{Department of Physics, Graduate School of Science, Kyoto University, Kyoto 606-8502, Japan}

\author{Yosuke Takasu}
\altaffiliation{}
 \affiliation{Department of Physics, Graduate School of Science, Kyoto University, Kyoto 606-8502, Japan}

\author{Shimpei Goto}
\affiliation{Institute for Liberal Arts, Institute of Science Tokyo, Ichikawa, Chiba 272-0827, Japan}

\author{Hironori Kazuta}
\affiliation{Department of Physics, Kindai University, Higashi-Osaka, Osaka 577-8502, Japan}

\author{Masaya Kunimi}
\affiliation{Department of Physics, Tokyo University of Science, Shinjuku, Tokyo 162-8601, Japan}

\author{Ippei Danshita}
\affiliation{Department of Physics, Kindai University, Higashi-Osaka, Osaka 577-8502, Japan}

\author{Yoshiro Takahashi}
\altaffiliation{}
 \affiliation{Department of Physics, Graduate School of Science, Kyoto University, Kyoto 606-8502, Japan}


\date{\today}

\begin{abstract}
While isolated quantum systems generally thermalize after long-time evolution, there are several exceptions defying thermalization. A notable mechanism of such nonergodicity is the Hilbert space fragmentation (HSF), where the Hamiltonian matrix splits into an exponentially large number of sectors due to the presence of nontrivial conserved quantities. Using ultracold gases, here we experimentally investigate the one-dimensional Bose-Hubbard system with neither disorder nor tilt potential, which has been predicted to exhibit HSF caused by a strong interatomic interaction. Specifically, we analyze far-from-equilibrium  dynamics starting from a charge-density wave of doublons (atoms in doubly occupied sites) in a singlon and doublon-resolved manner to reveal a slowing-down of the relaxation in a strongly interacting regime. We find that the numbers of singlons and doublons are conserved during the dynamics, indicating HSF as a mechanism of the observed slow relaxation. Our results provide an experimental confirmation of the conserved quantities responsible for HSF.
\end{abstract}


\maketitle


\section{Introduction}
The problem of quantum thermalization, i.e., how isolated quantum many-body systems which undergo the reversible unitary time evolution can reach thermal equilibrium states, lies at the heart of modern quantum statistical physics and is of considerable recent interest.
As a mechanism of the thermalization, the eigenstate thermalization hypothesis (ETH) is known, which states that expectation values of physical quantities for eigenstates of a quantum many-body system coincide with that of the microcanonical ensemble in the corresponding energy \cite{PhysRevA.43.2046,PhysRevE.50.888,rigol2008thermalization,deutsch2018eigenstate}.
Especially, when a quantum many-body system satisfies the strong version of the ETH, where all of the eigenstates satisfy the ETH, the long-time average of the physical quantity coincides with the microcanonical average, that is, the system thermalizes \cite{d2016quantum,mori2018thermalization}.

The remarkable progress of artificial quantum systems, such as ultracold gases, Rydberg atom arrays, trapped ions, and superconducting qubits, has enabled indispensable studies on the problem of the thermalization from both experimental and theoretical sides.
In better understanding of the mechanisms of the thermalization, the investigation of nonergodic systems, which do not show thermalization, is important. Illustrative examples of nonergodic systems include integrable \cite{kinoshita2006quantum,PhysRevLett.98.050405,PhysRevLett.103.100403,PhysRevA.80.053607,PhysRevLett.106.140405,vidmar2016generalized} and many-body localized (MBL) systems \cite{PhysRevLett.95.206603,basko2006metal,PhysRevB.90.174202,altman2015universal,nandkishore2015many,schreiber2015observation,choi2016exploring,PhysRevA.95.021601,sierant2017many,PhysRevLett.119.260401,sierant2018many,alet2018many,RevModPhys.91.021001,PhysRevLett.122.170403,sierant2025many}, where thermalization is prevented due to the presence of an extensive number of conserved quantities and the strong disorder potential, respectively.
Furthermore, the recent findings of another type of nonergodic systems show the diversity of origins of nonergodic behavior, as exemplified by quantum many-body scar \cite{bernien2017probing,PhysRevLett.124.160604,serbyn2021quantum,bluvstein2021controlling,Papić2022,moudgalya2022quantum,chandran2023quantum,su2023observation,kaneko2024quantum,PhysRevResearch.6.043259}, Stark MBL \cite{PhysRevLett.122.040606,van2019bloch,yao2021nonergodic,morong2021observation,PhysRevLett.127.240502} and Hilbert-space fragmentation (HSF) \cite{PhysRevX.10.011047,PhysRevB.101.174204,scherg2021observing,PhysRevLett.130.010201,PhysRevResearch.5.043239,wang2024exploring,PhysRevX.15.011035,Adler2024}.

In particular, HSF typically results from the presence of nontrivial conserved quantities, leading to strong kinetic constraints on the system.
Under such constraints, the Hilbert space is fragmented into an exponentially large number of disconnected subsectors (Krylov subsectors), where the dynamics are restricted to only a few of subsectors, causing the system not to thermalize.
Experimentally, the nonergodic dynamics due to the HSF have been observed, especially in the one-dimensional (1D) Fermi-Hubbard system \cite{scherg2021observing,PhysRevLett.130.010201} and the two-dimensional Bose-Hubbard system \cite{Adler2024}, where linear potential gradients (tilt potentials) are applied, and the total atom number and dipole moment are conserved.
Theoretically, a disorder-free 1D Bose-Hubbard model with no trapping potential \cite{carleo2012localization,PhysRevB.102.144302} and with a trapping potential \cite{yao2020many,PhysRevA.104.043322,yao2021nonergodic} has been studied.
Especially in Ref.~\cite{PhysRevA.104.043322}, it has been reported that the nonergodic behavior emerges in the strongly interacting regime when the dynamics start from a period-two charge-density wave (CDW) of doublons (atoms in doubly occupied sites) (see Fig.~\ref{Fig.1}A-B), while rapid relaxation occurs starting from that of singlons.
In this system, in addition to the total atom number, the numbers of doublons and singlons are emergent conserved quantities in the strongly interacting regime, which gives rise to strong HSF.

In this work, using ultracold Bose gases in optical lattices, we explore the role of interatomic interaction in the nonequilibrium dynamics following the sudden quench of the optical-lattice depth starting from a particular initial state of a period two CDW of doublons with a small portion of singlons in a 1D disorder-free Bose-Hubbard system with no tilt potential (See Fig.~\ref{Fig.1}A).
We experimentally observe a striking difference between the quench dynamics starting from this type of initial state and that from the period two CDW of singlons. 
The systematic measurement of the imbalance in a singlon- and doublon-resolved manner with varying the ratio of the interatomic interaction to the hopping energy reveals a slowing-down of the relaxation of the imbalance of the doublons in a strongly-interacting regime, in stark contrast with the behavior of the singlons which exhibit the considerable relaxation in all interaction strengths.
Importantly, we find that both of the numbers of singlons and doublons are conserved during the quench dynamics, indicating the HSF as a mechanism of the observed slow relaxation.
The interplay between the effects of parabolic potentials and inter-atomic interaction plays a role in the observed relaxation dynamics, which is reproduced by our theoretical calculation.
In particular, we reveal the enhanced slowing-down of the relaxation for doublons by partial removal of singlons that accelerate the equilibration of the imbalance through tunneling with no energy cost, which offers important insights into the quantum thermalization dynamics.
We note that, in contrast to the pioneering work \cite{winkler2006repulsively} where the atom number stability of the doublon itself, i.e., repulsively-bound pair, prepared in an approximately isolated form (filling of doublons is typically 0.3) is observed, the present work reveals the nonequilibrium dynamics for interacting many-body systems of the doublons by observing the atom-number imbalance between the odd and even sites.

\section{Results}
\noindent
{\bf Experimental setup}\\
We start with the preparation of a $^{174}$Yb Bose-Einstein condensate (BEC) by evaporative cooling with the total atom number of about $1.3\times 10^4$.
A Mott insulating state of unit filling is formed after the atom loading into a 3D optical lattice with a deep potential depth of 30$E_{\rm R}$, where $E_{\rm R}=\hbar^2k_{\rm L}^2/(2m)$ is the recoil energy of the optical lattice, $k_{\rm L}=2\pi/\lambda_{\text {short}}$ with $\lambda_{\text {short}}=532$~nm is the wave number of the laser for the optical lattice, and $\hbar$ is the Planck constant divided by 2$\pi$.
Here, the preparation of the initial state of period-two CDW of doublons, i.e., $\ket{\psi(0)}_{\text{CDW(d)}}=\ket{\cdots 2020 \cdots}$ state in 1D chains proceeds with the optical superlattice in the direction of 1D chains consisting of short ($\lambda_{\text{short}}=532$~nm) and long ($\lambda_{\text{long}}=1064$~nm) lattices (see Sec.~S.1
for details of the loading procedure). 
For representing the many-body state of a 1D chain with $M$ sites, here we use the Fock basis,
\begin{eqnarray}
|n_1\, n_2 \,\cdots\, n_M\rangle \equiv \bigotimes_{i=1}^M\ket{n_i}_i,
\end{eqnarray}
where $\ket{n_i}_i$ denotes the local Fock state at a site $i$ with an atom number $n_i$.
The central 1D chains of the atoms have a length of about 25.
We note that in a prepared initial state, 25-30$\%$ of total atoms are singly occupied, which is confirmed by the remaining fraction of the atoms after the irradiation of a photoassociation (PA) laser.
This is induced by the presence of low-density regions at the trap edges and the nonadiabaticity in the loading process (see Sec.~S.1 for the estimation of the singlon fraction in the initial state).

\begin{figure*}
  \includegraphics[width=17.2cm]{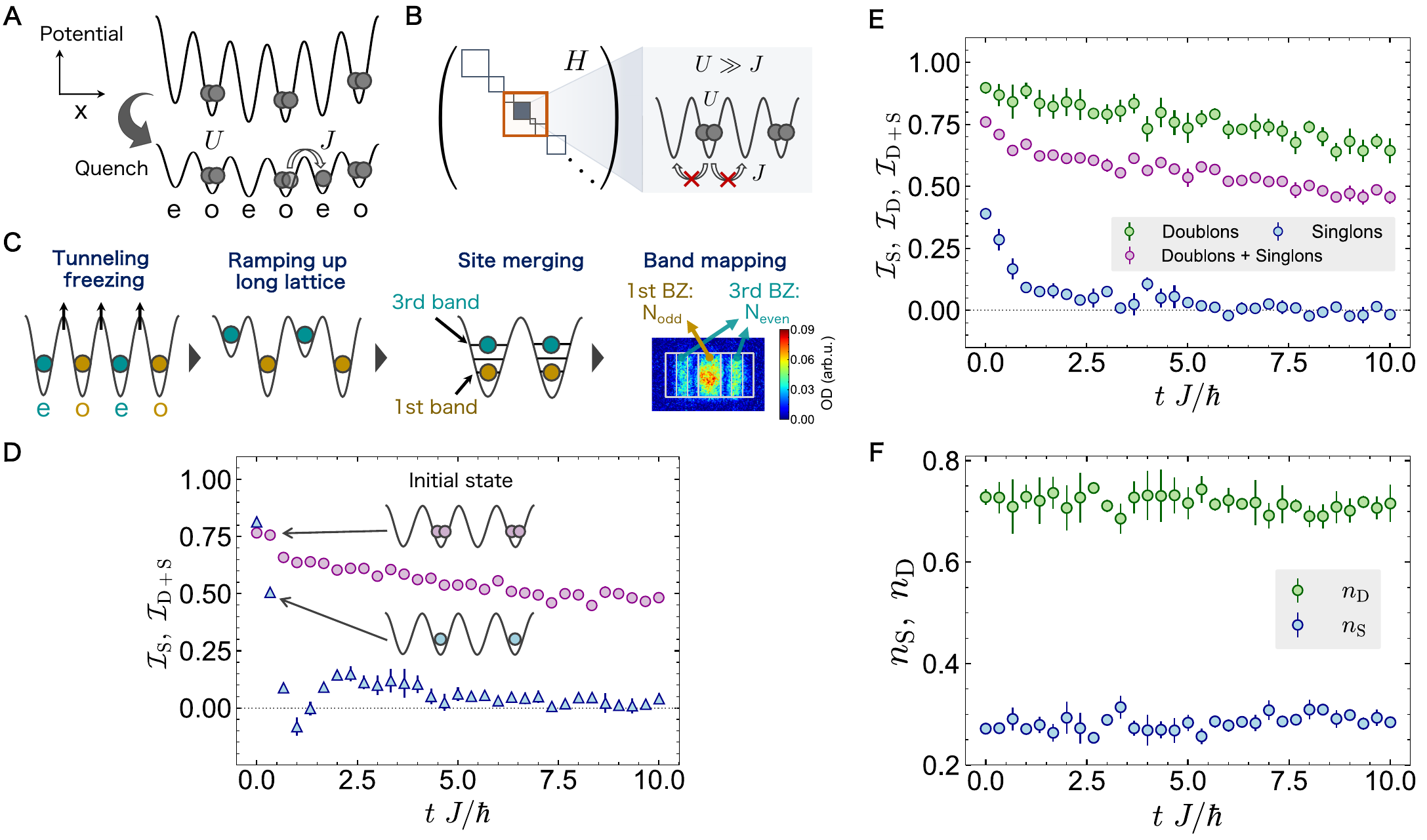}
  \caption{\label{Fig.1} {\bf Schematic of the experimental setup and illustration of observed quench dynamics.}
  ({\bf A}) Schematic of the 1D Bose-Hubbard model.
  ({\bf B}) Schematic illustration of HSF.
  The orange square shows a subspace characterized by the particle number.
  This subspace is further fragmented into an exponentially large number of subsectors (Krylov subsectors) under some kinetic constraints, where the breakup of doublons into singlons is suppressed for $U \gg J$.
  ({\bf C}) Schematic illustration of the sequence for a site-mapping technique to measure the atom number imbalance.
  ({\bf D}) Comparison of quench dynamics at $U/J=52$ starting from $\hat{\rho}_{\text{CDW(d,s)}}(0)$ (purple circle) and $\hat{\rho}_{\text{CDW(s)}}(0)$ (blue triangle) as a function of the normalized holding time with the tunneling time of $\hbar/J=2.6$~ms along the direction of the 1D chains.
  Error bars in $\mathcal{I}_{\rm D+S}$ and $\mathcal{I}_{\rm S}$ show the standard deviation of three and five independent scans, respectively.
  ({\bf E}) Typical imbalance dynamics.
  ({\bf F}) Typical dynamics of atom number fractions of doublons~($n_{\mathrm{D}}$, green) and singlons~ ($n_{\mathrm{S}}$, blue) as a function of the normalized holding time with the tunneling time of $\hbar/J=2.1$~ms along the direction of the 1D chains.
  In (E) and (F), all data are obtained at $U/J=67$.
  Error bars for $\mathcal{I}_{\rm D+S}$ and $\mathcal{I}_{\rm S}$ in (E) show the standard deviation of three independent scans, and those for $\mathcal{I}_{\rm D}$ in (E) and $n_{\mathrm{S/D}}$ in (F) show the standard deviation calculated by the error propagation formula.
  }
\end{figure*}

After the initial state preparation, we perform a sudden quench by rapidly decreasing the potential depth to initiate the dynamics along the 1D chains (Fig.~\ref{Fig.1}A. See also 
Materials and Methods for the details of the quench procedure).
The dynamics during the hold time in the 1D chains is described by the 1D Bose-Hubbard model, and the Hamiltonian is given by
\begin{equation}\label{eq:Hamiltonian} 
    \hat{H}=-J\sum_{\langle i,j \rangle} \hat{a}_i^{\dagger}\hat{a}_j + \frac{U}{2}\sum_{i=1}^{M} \hat{n}_i(\hat{n}_i-1) + \sum_{i=1}^{M} V_i \hat{n}_i,
\end{equation}
where $\hat{a}_i(\hat{a}_i^{\dagger})$ is the annihilation (creation) operator of a boson at a site $i$; $J$ is the tunneling amplitude between nearest-neighbor sites $\langle i,j \rangle$;
$\hat{n}_i\equiv\hat{a}_i^{\dagger}\hat{a}_i$ is the number operator at a site $i$; $U$ is the on-site interaction strength; $V_i\equiv\Omega [i-(M+1)/2]^2$ is the parabolic potential; and $\Omega$ is the strength of the parabolic potential \cite{PhysRevA.104.043322}.
Here, we note that there is no tilt potential in Eq.~(\ref{eq:Hamiltonian}), different from previous experimental studies of nonergodic dynamics due to the HSF \cite{scherg2021observing,PhysRevLett.130.010201,Adler2024} and the many-body scarring \cite{su2023observation}.
Note that the relative strength of the parabolic potential $\Omega$ to the tunneling amplitude $J$ depends on the lattice depth (See Tables~S1
and S2 for specific values of $\Omega/J$),
and
$\Omega$ is small compared with hopping $J$ of the singlon but not with that of the doublon. 

In order to detect the atom distribution after the quench, we freeze the dynamics by rapidly ramping up the potential depth along the 1D chains.
In this work, we focus on two physical quantities that characterize the atom distribution: atom-number imbalance $\mathcal{I}(t)$ between the odd and even sites, and doublon- and singlon-resolved atom-number fraction ($n_{\mathrm{D}}, n_{\mathrm{S}}$).
The imbalance $\mathcal{I}(t)$ is defined as 
\begin{equation} \label{eq:imb_def}
    \mathcal{I}(t) \equiv \frac{N_{\text{odd}}(t)-N_{\text{even}}(t)}{N_{\text{odd}}(t)+N_{\text{even}}(t)},
\end{equation}
which becomes zero for thermalized states, where $N_{\text{odd}}(t)$ and $N_{\text{even}}(t)$ are the atom numbers in the odd and even sites, respectively.
The measurement of the imbalance is performed by a site-mapping technique; i.e., mapping of the even and odd sites to the 1st and 3rd bands, respectively, followed by band mapping \cite{PhysRevA.73.033605,folling2007direct,scherg2021observing,PhysRevLett.130.010201} (see Fig.~\ref{Fig.1}C and Materials and Methods 
for the details of the imbalance measurement).
Note that, as we mention in the state preparation, the prepared state is not a pure state $\ket{\psi(0)}_{\text{CDW(d)}}$, but a mixed state $\hat{\rho}_{\text{CDW(d,s)}}(0)$, which involves singlons with 25-30$\%$ of total atoms.
By exploiting the PA resonance which selectively excites and thus removes the doublons, we achieve measurements of the imbalance in a singlon- and doublon-resolved manner \cite{folling2007direct,PhysRevLett.130.010201}.
Specifically, we measure the imbalance with and without the PA laser irradiation just after the freezing of the dynamics and before the site-mapping, where the imbalance of singlons ($\mathcal{I}_{\text{\ w/ PA}}=\mathcal{I}_{\mathrm{S}}$) and that of both doublons and singlons ($\mathcal{I}_{\text{\ w/o PA}}$) are obtained, respectively.
Then, we obtain the imbalance of doublons ($\mathcal{I}_{\mathrm{D}}$) from the former two measured values as follows:
\begin{equation} \label{eq:I_D_def}
    \mathcal{I_{\mathrm{D}}}=\frac{\mathcal{I}_{\text{\ w/o PA}} \cdot N_{\text{w/o PA}} - \mathcal{I}_{\text{\ w/ PA}} \cdot N_{\text{w/ PA}}}{N_{\text{w/o PA}}-N_{\text{w/ PA}}},
\end{equation}
where $N_{\text{w/ PA}}$ ($N_{\text{w/o PA}}$) is the atom number measured with (without) PA laser irradiation.
The doublon- and singlon-resolved atom number fraction, $n_{\mathrm{D}}$ and $n_{\mathrm{S}}$, are obtained with the atom numbers measured with and without the PA laser irradiation, $N_{\text{w/ PA}}$ and $N_{\text{w/o PA}}$, as follows:
\begin{align}
    n_{\mathrm{S}} &= N_{\text{w/ PA}}\ /\ N_{\text{w/o PA}},\\
    n_{\mathrm{D}} &= (N_{\text{w/o PA}}-N_{\text{w/ PA}})\ /\ N_{\text{w/o PA}}.
\end{align}
Note that these atom number fractions are important to characterize the nonequilibrium dynamics in the sense that they are emergent conserved quantities in the strongly interacting regime of the Bose-Hubbard chain, leading to the HSF~\cite{PhysRevA.104.043322}.\\

\noindent
{\bf Typical quench dynamics: slow imbalance relaxation, and doublon- and singlon-fraction conservation}

\begin{figure*}
  \includegraphics[width=17.2cm]{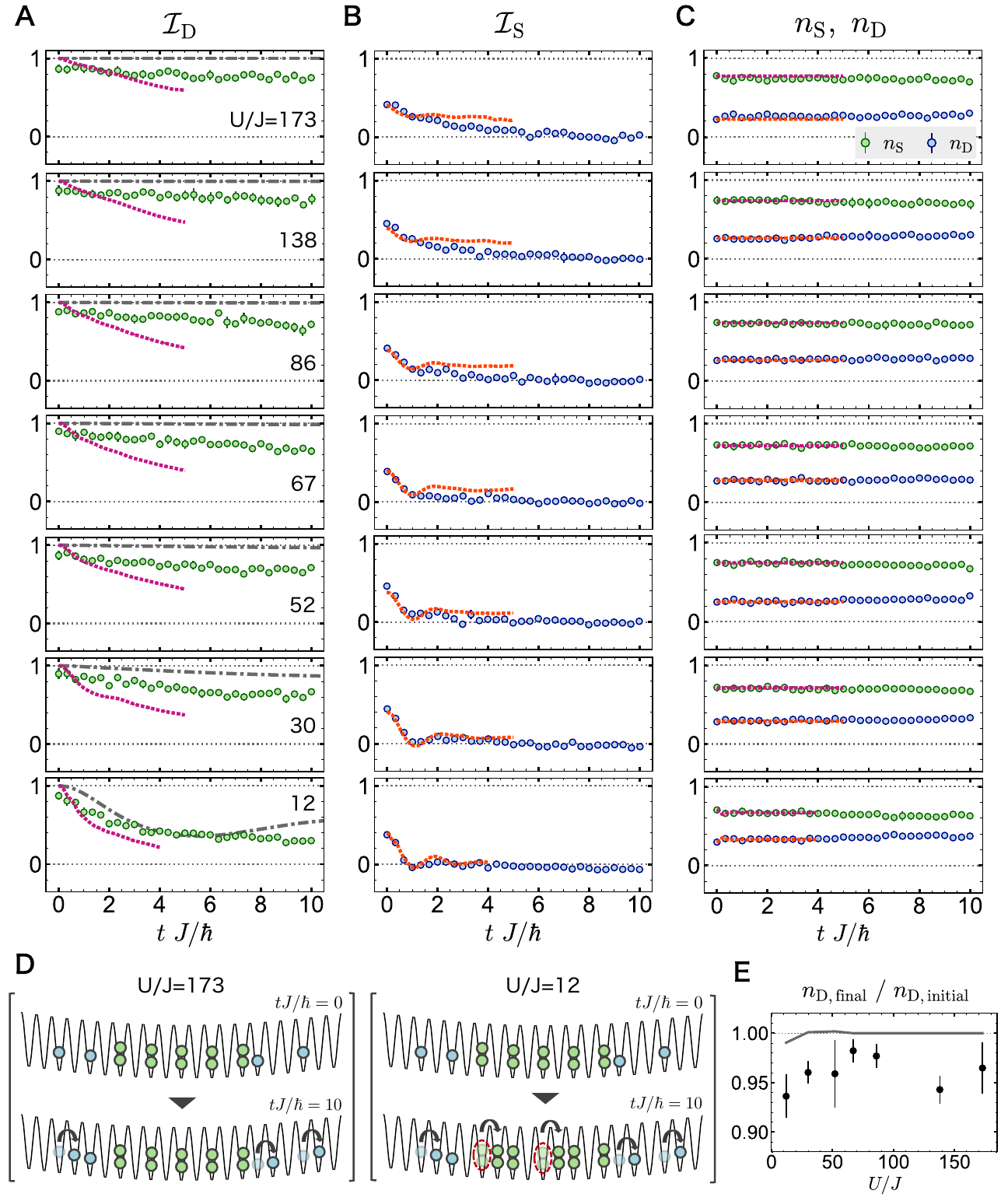}
  \caption{\label{Fig.2} {\bf Dependence of quench dynamics on $U/J$.} ({\bf A},{\bf B}) Imbalance dynamics of (A)~doublons ($\mathcal{I}_{\mathrm{D}}$), (B)~singlons ($\mathcal{I}_{\mathrm{S}}$), and ({\bf C})~atom number fractions of doublons~($n_{\mathrm{D}}$, green) and singlons~($n_{\mathrm{S}}$, blue) on $U/J$.
  Error bars in (A)-(C) representing the standard deviation are mostly smaller than symbols.
  In (A)-(C), dotted curves show the calculated values.
  In (A), dash-dotted curves show the calculated values for hardcore bosons as a reference (see Fig.~3 for details).
  Error bars of calculated values representing the statistical error are smaller than the dots of the curves.
  ({\bf D}) Schematic illustration of tunneling dynamics of doublons and singlons in a lattice for relatively small (right) and large (left) $U/J$ cases.
  ({\bf E}) Change of the doublon fraction during the hold time.
  Here, the ratio of the averaged doublon fraction in $tJ/\hbar=9$-$10$ ($n_{\mathrm{D},\text{final}}$) to that in $tJ/\hbar=0$-$1$ ($n_{\mathrm{D},\text{initial}}$) is plotted.
  Error bars show the standard deviation, which is calculated by the error propagation formula with the standard deviation of the averaged doublon fractions $n_{\mathrm{D},\text{initial}}$ and $n_{\mathrm{D},\text{final}}$.
  The solid curve shows the ratio of the calculated averaged doublon fraction in $tJ/\hbar=4$-$5$ ($tJ/\hbar=3$-$4$ only in the case of $U/J=12$) ($n_{\mathrm{D},\text{final}}$) to that in $tJ/\hbar=0$-$1$ ($n_{\mathrm{D},\text{initial}}$).
  }
\end{figure*}
\noindent
We first show the comparison of the relaxation behavior obtained at $U/J=67$ starting from the $\hat{\rho}_{\text{CDW(d,s)}}(0)$ and $\hat{\rho}_{\text{CDW(s)}}(0)$ states in Fig.~\ref{Fig.1}D (see Sec.~S.2
for the details of the quench dynamics starting from the $\hat{\rho}_{\text{CDW(s)}}(0)$ state). Here, $\hat{\rho}_{\text{CDW(s)}}(0)$ denotes an initial mixed state that is in reality created when we try to ideally prepare
\begin{eqnarray}
\ket{\psi(0)}_{\text{CDW(s)}}=\ket{\cdots 1010 \cdots}.
\end{eqnarray}
In Fig.~\ref{Fig.1}D, the imbalance $\mathcal{I}$ shows slow relaxation and remains far from zero at least up to ten times of the tunneling time along the 1D chains when the initial state is $\hat{\rho}_{\text{CDW(d,s)}}(0)$, while the imbalance shows rapid relaxation and an oscillating behavior across zero when the initial state is $\hat{\rho}_{\text{CDW(s)}}(0)$.
These different relaxation behaviors are theoretically discussed in Ref.~\cite{PhysRevA.104.043322}, where the nonergodic dynamics is expected when the initial state is $\ket{\psi(0)}_{\text{CDW(d)}}=\ket{\cdots 2020 \cdots}$.
Note that although the relaxation behavior in the 1D Bose-Hubbard system when the initial state is approximately $\ket{\cdots 1010 \cdots}$ has already been reported in Ref.~\cite{trotzky2012probing}, we revisit this situation to show a clear difference in the quench dynamics depending on the initial states.

In the following, for the quantitative study, we focus on the quench dynamics starting from the $\hat{\rho}_{\text{CDW(d,s)}}(0)$ state.
In Fig.~\ref{Fig.1}E, we show the typical quench dynamics of the imbalances of both doublons and singlons (`Doublons$+$Singlons', $\mathcal{I}_{\mathrm{D+S}}$), singlons (`Singlons', $\mathcal{I}_{\mathrm{S}}$), and doublons (`Doublons', $\mathcal{I}_{\mathrm{D}}$) obtained from the former two by Eq.~(\ref{eq:I_D_def}), where $U/J=67$.
Here, the imbalances of doublons and singlons show slow and rapid relaxation, respectively.
Note that tunneling time along the direction perpendicular to the 1D chains is $\hbar/(4J_{\perp})=92$~ms, about 3.5~times longer than the maximum measurement time.
In addition to the successful observation of the slow relaxation behavior for doublons, we observe that both of the atom number fractions of doublons ($n_{\mathrm{D}}$) and singlons ($n_{\mathrm{S}}$) are almost conserved, as shown in Fig.~\ref{Fig.1}F.
This reveals that the atom numbers of both doublons and singlons serve as approximate conserved quantities, which is a key finding in the sense that this strongly supports the occurrence of HSF in our system, where the dissociation of doublons due to single particle tunnelings is suppressed for $U \gg J$ due to the energy mismatch (see Fig.~\ref{Fig.1}B).
We note that the difference between the mechanism of HSF in the tilted systems of the previous studies and that in our system with no tilt is in nontrivial emergent conserved quantities that cause HSF.
This conserved quantity is the dipole moment in addition to the total number of particles in the previous study with tilt, while in our system, it is the number of doublons and singlons in a strongly interacting regime in addition to the total number of particles.
We find that the magnitude of $\Delta/U$ in this study is much smaller than one, where $\Delta$ is the maximum energy offset between neighboring sites in the parabolic trap in our system (see Tables~S1 and S2 for specific values of $\Delta/U$).
Therefore, our study is not in the region of $\Delta/U\gtrsim1$ where HSF is caused by tilt potential as in previous studies.
From these discussions, we also note that the HSF of our system is not sensitive to the system size, since the kinetic constraint comes mainly from the strong interaction rather than the trapping potential \cite{PhysRevA.104.043322}.\\

\noindent
{\bf Dependency of quench dynamics on $U/J$}\\
In order to elucidate the role of interatomic interaction as well as the interplay with the parabolic potential in the nonequilibrium dynamics, we investigate the dependency of the quench dynamics on the interaction strength $U/J$ and the parabolic trap strength $\Omega/J$ by changing the potential depth along the 1D chains after the quench, shown in Fig.~\ref{Fig.2}A-C.
Here, in Fig.~\ref{Fig.2}A and \ref{Fig.2}B, we observe an overall tendency of slower relaxation of the imbalance of doublons for larger $U/J$, while the imbalance of singlons shows fast relaxation roughly independent of $U/J$.
At the same time, in Fig.~\ref{Fig.2}C, we observe that the atom number fractions of both doublons and singlons are almost conserved regardless of the value of $U/J$ (see also Fig.~\ref{Fig.2}E).
These results can be understood intuitively as follows:
Recall that the dissociation of a doublon into two singlons is approximately forbidden, as assured by the conservation of the fractions.
In the cases of relatively small $U/J$, not only singlons can tunnel to adjacent vacant sites with the tunneling rate of $J$, but also doublons via the process of the second-order perturbation with the effective tunneling rate of $J_{\rm eff}=2J^2/U$, as depicted in Fig.~\ref{Fig.2}D~(right), on the one hand.
In fact, the time constant of the correlated tunneling, $\hbar/(2J_{\rm eff})$ is $3\hbar/J$ for the smallest $U/J$ case ($U/J=12$), where the factor of $2$ in front of $J_{\rm eff}$ means the number of nearest-neighboring sites along the chain direction.
In the cases of relatively large $U/J$, on the other hand, the rate of the effective hopping of the doublon becomes quite small, corresponding to about 1/5-time hopping event during the maximum holding time, for largest $U/J$ case ($U/J=173$), and thus only singlon tunneling is allowed, as in Fig.~\ref{Fig.2}D~(left).
Note that the effect of the parabolic trap also contributes to the suppression of the doublon tunneling, while such an effect is considerably weaker for singlons.\\

\noindent
{\bf Comparison between experiments and numerical calculations}\\
In order to provide a reference to be compared with the experimental results, we compute the real-time dynamics of the 1D Bose-Hubbard model of Eq.~(\ref{eq:Hamiltonian}) by using the time-evolving block decimation (TEBD) method~\cite{vidal2004}, which is based on matrix-product state (MPS) representation of a quantum many-body state~\cite{schollwock2011}. For TEBD, we use a second-order Suzuki-Trotter decomposition of the time evolution operator and optimally choose the time step $\Delta t$ within the range $0.005\leq \Delta t \,J/\hbar \leq 0.025$, depending on the interaction strength $U/J$. We set the maximal bond dimensions of MPS to be 2000 such that the time duration accessible with the numerical calculations is as short as or even shorter than the half of that in the experiments. In Sec.~S.5,
we elaborate on our theoretical protocol of initial-state preparation imitating the experimental situation.
In Fig.~\ref{Fig.2}A-\ref{Fig.2}C, we show the direct comparison of measured and calculated values of (A)~doublon- and (B)~singlon-resolved imbalance ($\mathcal{I}_{\mathrm{D}}$ and $\mathcal{I}_{\mathrm{S}}$) and (C)~doublon and singlon fractions ($n_{\mathrm{D}}$ and $n_{\mathrm{S}}$) for several values of $U/J$.
The quantitative disagreement between the experiment and simulations for ${\cal I}_{D}$ in Fig.~2 may come from the limitation of the simulation to completely reflect the entire adiabatic and nonadiabatic process of the initial states preparation in the experiment, which may well be in the non-thermal distribution of the singlons.
One can see, however, that the experimental data are within the range of the two simulation results of the case where the thermal initial state is given in the manner described in Sec.~S.5 and that where the initial state does not include the singlons at all as another extreme case.\\

\noindent
{\bf Competition among doublon-doublon interactions, doublon-singlon interactions, and a parabolic trap in the imbalance dynamics}\\
\begin{figure*}
  \includegraphics[width=17.2cm]{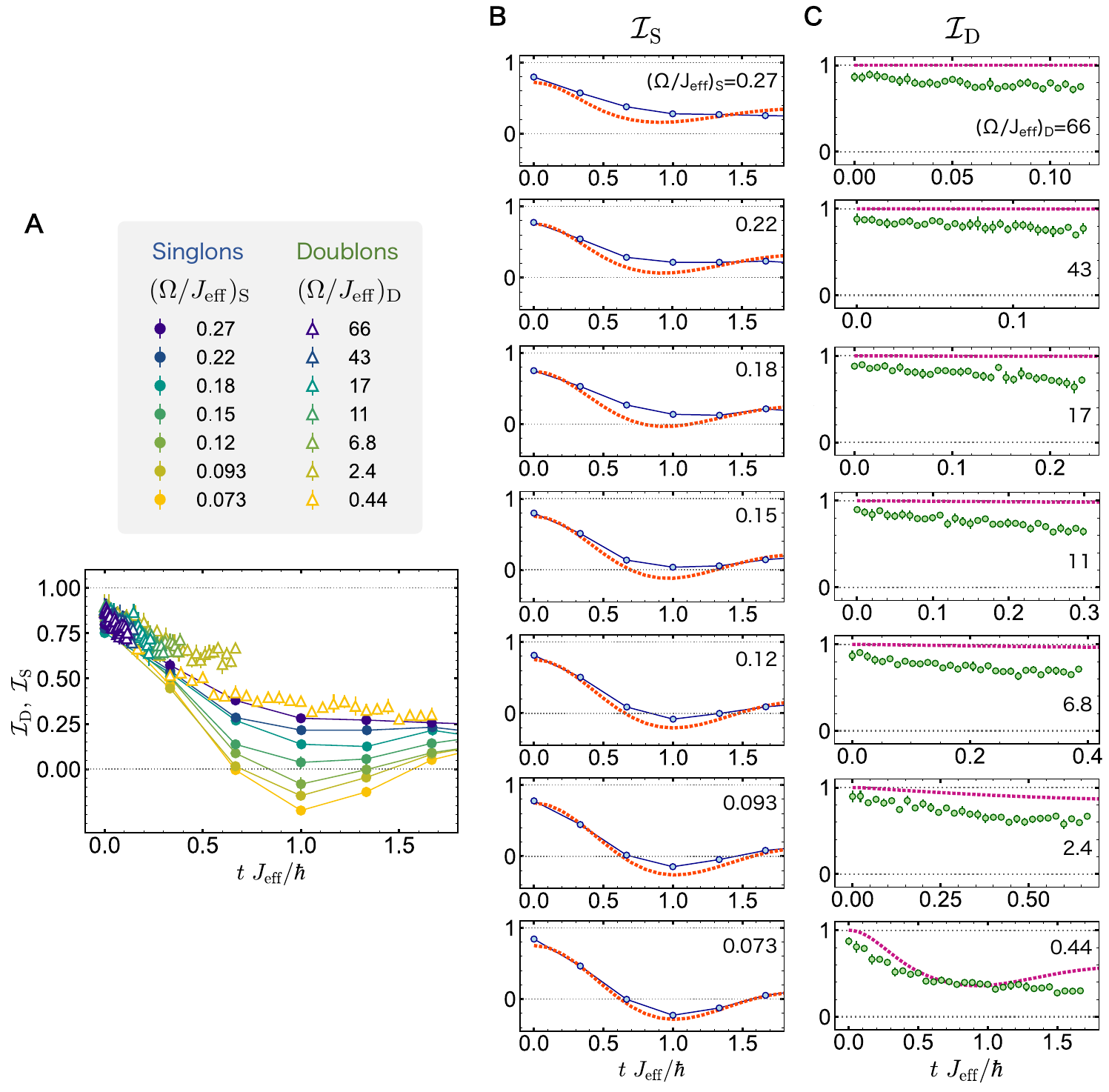}
  \caption{\label{Fig.3} {\bf Comparison of the measured imbalance dynamics of singlons starting from $\hat{\rho}_{\rm CDW(s)}(0)$ and those of doublons starting from $\hat{\rho}_{\rm CDW(d,s)}(0)$ with the numerical calculations.}
  ({\bf A}) Systematically measured imbalance dynamics of singlons (solid circles) and doublons (open triangles) for various $(\Omega/J_{\rm eff})_{\rm{D}}$ and $(\Omega/J_{\rm eff})_{\rm{S}}$.
  The horizontal axis represents the normalized time by the effective tunneling amplitude $J_{\rm{eff}}$, where $J_{\rm{eff}}=J$ for singlons and $J_{\rm{eff}}=2J^2/U$ for doublons.
  ({\bf B}), ({\bf C}) Comparison of measured imbalance dynamics of (B) singlons and (C) doublons with the calculation for singlons (dotted curves) representing the case of non-interacting hardcore bosons. In the calculations of (B), the initial state is a mixed state $\hat{\rho}_{\rm CDW(s)}(0)$ while in (A), it is a pure state $\ket{\psi}_{\rm CDW(s)}(0)$.
  Error bars in (A)-(C) representing the standard deviation are mostly smaller than symbols.
  }
\end{figure*}
\begin{figure*}
  \includegraphics[width=17.2cm]{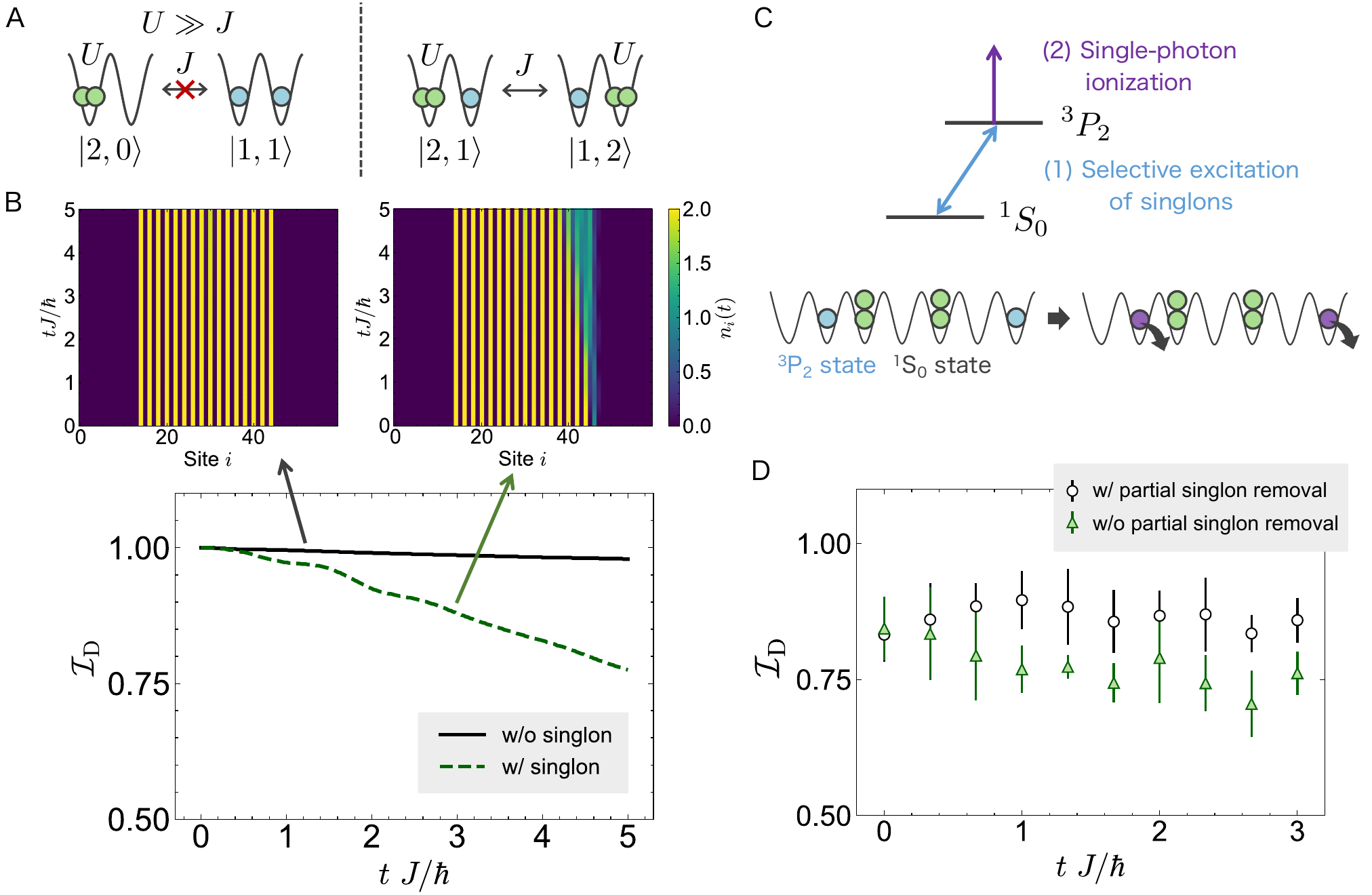}
  \caption{\label{Fig.4} {\bf Effect of singlons on the imbalance dynamics.}
  ({\bf A}) Schematic illustration of possible tunneling processes involving singlons between nearest-neighbor sites.
  ({\bf B}) Calculated time evolution of the local density in the 1D Bose Hubbard model (Eq.~(\ref{eq:Hamiltonian})) for $U/J=52$, starting from $\ket{\psi(0)}_{\text{CDW(d)}}=\ket{\cdots 2020 \cdots}$ state (upper left) and $\hat{a}^{\dagger}_{\rm edge}\ket{\psi(0)}_{\text{CDW(d)}}$ state (upper right), where one singlon is placed at the right edge of the trap in this case.
  (lower) Imbalance dynamics of doublons ($\mathcal{I}_{\mathrm{D}}$) obtained from the data in (upper).
  ({\bf C}) Schematic illustration of the removal procedure of only singlons from the initial $\hat{\rho}_{\text{CDW(d,s)}}(0)$ state, and the relevant energy levels.
  ({\bf D}) Imbalance dynamics of doublons after partial removal of singlons in the initial $\hat{\rho}_{\text{CDW(d,s)}}(0)$ state.
  Here, the plots of open circle and solid triangle show the results with ($n_{\mathrm{S}}=0.24$) and without ($n_{\mathrm{S}}=0.29$) partial removal of singlons, respectively.
  Note that the loss of doublons in this measurement is negligibly small within the uncertainty.
  Both data are obtained at $U/J=52$.
  All error bars show the standard deviation.
  }
\end{figure*}
\noindent
While we see the agreement between the experimental data and theoretical calculations, here we discuss how much extent the experimental results can be understood in an intuitive manner and clarify the 
nontrivial feature beyond simple interpretations.
Specifically, we first see whether we can understand in a unified manner the dynamics of the singlons and doublons by regarding both of them as simple non-interacting hardcore bosons with the corresponding effective tunneling amplitude $J_{\rm{eff}}=J$ for the singlons and $J_{\rm{eff}}=2J^2/U$ for the doublons placed in a parabolic trap with the trap strength $\Omega$, thus characterized only by $(\Omega/J_{\rm eff})_{\rm S}=\Omega/J$ $(\Omega/J_{\rm eff})_{\rm D}=(\Omega/J_{\rm eff})_{\rm S}\cdot (U/J)$, respectively.
In Fig.~\ref{Fig.3}, we show a comparison of the experimental results of the imbalance dynamics for the singlons for the initial $\hat{\rho}_{\text{CDW(s)}}(0)$ state and those for the doublon for the initial $\hat{\rho}_{\text{CDW(d,s)}}(0)$ state as a function of the time normalized by the effective tunneling amplitude $J_{\rm{eff}}$. (see also Sec.~S.2 
for the quench dynamics starting from a $\hat{\rho}_{\text{CDW(s)}}(0)$ state).
Note that, due to the large difference in the amplitude of $J_{\rm{eff}}$ between the singlons and doublons, the parameter regions of $(\Omega/J_{\rm eff})_{\rm S}$ and $(\Omega/J_{\rm eff})_{\rm D}$ have no overlap.
Here, instead of the direct comparison of the experimental data, as shown in Fig.~\ref{Fig.3}C, we focus on the comparison of the experimental results for the doublons with the theoretical calculations for the singlons with the parameter region of $(\Omega/J_{\rm eff})_{\rm S}$ beyond 0.44 in accord with that for the doublons.
Since the calculation for the singlons is rather straightforward, and indeed we see that the calculations well reproduce the experimental results for $(\Omega/J_{\rm eff})_{\rm S}$ below 0.23, as also shown in  Fig.~S3, 
we well expect the validity of the calculations for the singlons also for $(\Omega/J_{\rm eff})_{\rm S}$ beyond 0.44, which should be compared with the experimental results for the doublons in the corresponding region of $(\Omega/J_{\rm eff})_{\rm D}$.
While the slow, non-oscillatory relaxation behavior is commonly observed both for the experiments and the calculations in Fig.~\ref{Fig.3}C, there is an overall discrepancy in that the experiments for the doublons show faster relaxation than the calculations.
Note the effective parabolic trap strengths of $(\Omega/J_{\rm eff})_{\rm S}$ in the calculation and $(\Omega/J_{\rm eff})_{\rm D}$ in the experiments are the same in this comparison.
The result of the comparison thus indicates that the doublons in this experiment cannot be simply interpreted as the non-interacting hardcore bosons, and we need to take interaction effects into consideration for full understanding of the observed behaviors.

First, we consider an effective nearest-neighbor doublon-doublon interaction, which is described in the following effective Hamiltonian for $U/J\gg 1$ \cite{carleo2012localization}:
\begin{equation}
  H^{\rm eff}=\frac{2J^2}{U}\sum_{\<i,j\>}\left[ -8\hat{S}_i^{z}\hat{S}_j^{z}+\left( \hat{S}_i^+\hat{S}_j^- + \hat{S}_i^-\hat{S}_j^+ \right) \right],
\end{equation}
where the doublon and holon (empty site) are associated to a fictitious spin up $\ket{\uparrow_i}$ and down $\ket{\downarrow_i}$, respectively, and $\hat{S}_i^z \equiv (\ket{\uparrow_i}\bra{\uparrow_i}-\ket{\downarrow_i}\bra{\downarrow_i})/2$ and $\hat{S}_i^+ \equiv \ket{\uparrow_i}\bra{\downarrow_i}=(\hat{S}_i^-)^{\dagger}$ are the spin-1/2 operators.
Here, the first term on the right-hand side corresponds to the doublon-doublon interaction, which takes the same order of magnitude as that of the second term for the doublon hopping.
This doublon-doublon interaction introduces the energy difference between the state of doublons in the CDW and a state that possesses doublons occupying in the nearest-neghboring sites, causing the slower relaxation of the initially prepared CDW of the doublons compared with the case of non-interacting hardcore bosons, in contrast to the observation in Fig.~\ref{Fig.3}C.

Next, we consider the interaction between the doublons and the singlons involved in the prepared initial $\hat{\rho}_{\text{CDW(d,s)}}(0)$ state. See the situations depicted in Fig.~\ref{Fig.4}A.
While the tunneling process of $\ket{2,0} \leftrightarrow \ket{1,1}$ is suppressed for $U \gg J$ as shown in Fig.~\ref{Fig.4}A~(left), that of $\ket{2,1} \leftrightarrow \ket{1,2}$, shown in Fig.~\ref{Fig.4}A~(right), has no energy cost, causing the faster relaxation of the initially prepared CDW of the doublons, consistent with the observation in Fig.~\ref{Fig.3}C.
This behavior is also numerically confirmed as shown in Fig.~\ref{Fig.4}B, where we depict the time evolution of the atom density profile in a single Bose-Hubbard chain starting from the $\ket{\psi(0)}_{\text{CDW(d)}}=\ket{\cdots 2020 \cdots}$ state (left) and $\hat{a}_{\rm edge}^{\dagger}\ket{\psi(0)}_{\text{CDW(d)}}=\ket{\cdots 2020 \cdots}$ (right). Here $\hat{a}_{\rm edge}^{\dagger}$ creates an atom at a site $i_{\rm edge}+2$, where $i_{\rm edge}$ denotes the right-edge site of the initial doublon array.
In the latter case, one singlon initially placed at the right edge of the trap moves to the trap center and disturbs $\cdots 2020 \cdots$ configuration, which leads to the substantial decay of $\mathcal{I}_{\rm D}(t)$ as seen in the bottom panel of Fig.~\ref{Fig.4}B.
From these discussions, we attribute the disagreement between the experiment and the simulations based on the simple hardcore boson picture for ${\cal I}_D$ in Fig.~3 to the interaction between singlons partially present in the initial state and doublons, as shown in Fig.~\ref{Fig.4}A(right), highlighting the unique feature of the singlon-induced equilibration mechanism.

In order to experimentally confirm the effect of initially involved singlons on the quench dynamics, we prepare an initial $\hat{\rho}_{\text{CDW(d,s)}}(0)$ state with a smaller singlon fraction by partially removing singlons with a selective ionization method, as depicted in Fig.~\ref{Fig.4}C (see also Sec.~S.4 
for the details of the removal scheme of singlons).
In Fig.~\ref{Fig.4}D, we show the imbalance dynamics of doublons starting from a $\hat{\rho}_{\text{CDW(d,s)}}(0)$ state with and without the partial singlon removal as the open circle and the solid triangle, respectively.
Importantly, we observe the slower relaxation of imbalance of doublons in the case of a smaller singlon fraction in the initial state.
In this way, we confirm the 
intriguing role of initially involved singlons in the quench dynamics, which qualitatively agrees with the numerical result.

\section{Discussion}
In summary, we experimentally studied nonequilibrium dynamics of 1D Bose gases in optical lattices, which is well described as 2D arrays of independent Bose-Hubbard chains, after a sudden quench of the optical lattice depth. Starting from an initial period-two CDW state of doublons slightly mixed with singlons,
we successfully observed 
slow relaxation in an imbalance measurement.
We found that the numbers of both doublons and singlons are almost conserved during the dynamics, which strongly supports the occurrence of HSF.
Systematic measurements of quench dynamics for various $U/J$ and the quantitative comparison of measured and calculated dynamics were also performed.
In addition, we revealed the effect of singlons involved in the initial $\ket{\cdots 2020 \cdots}$ state on the quench dynamics.

It is the time constant of hopping between 1D tubes that limits the observed time scale in our experiment.
By limiting ourselves to a shorter observation time than this time constant, we safely provide a clear discussion by separating unnecessary factors that could cause deviations from the 1D system we focus on in this study.
Importantly, the presence of singlons in the initial state of a period-two CDW of doublons causes the interaction between doublon and singlon in the dynamics, which occurs on the time scale of the bare tunneling time.
The doublon-singlon interaction results in richer dynamics already in the time window of less than 10 tunneling times, observed as a deviation from the non-interacting hardcore boson picture in doublon dynamics.
Our system has enough room for improvement for a longer observation time.
If we can further deepen the optical lattice in the direction perpendicular to the tube in our system, we will be able to see dynamics for longer times.
Here, since our lattice beams are not blue-detuned, we do not have to worry about doublon lifetime as in Ref.~\cite{PhysRevLett.130.010201}.

The successful preparation of arrays of dense doublons, i.e., $\ket{\cdots 2020 \cdots}$ state, and observations of dynamics of interacting doublons with a long lifetime open the possibility to study systems consisting of strongly-interacting composite particles with ultracold gases. For instance, it is interesting to examine whether some exotic quantum many-body states formed by such composites can be experimentally realized, such as the pair superfluids of bosons~\cite{daley2009}, the $\eta$-pairing state of two-~\cite{kantian2009} or multi-~\cite{PhysRevResearch.6.043259} component fermions, and quantum many-body scar states of bosons with a three-body constraint~\cite{kaneko2024quantum}.
More specifically, these states are expected to be prepared by adiabatically increasing the hopping from zero and controlling the energy offset of the double wells when one starts from the $|\ldots2020\ldots\rangle$ state.
Also, for the tilted 1D Bose-Hubbard system, previously studied in Ref.~\cite{su2023observation}, we expect that our measurement method of dynamics of singlons and doublons in a singlon- and doublon-resolved manner may provide further insights into quench dynamics in that system.

\section{MATERIALS AND METHODS}
\noindent
{\bf Sudden quench of the lattice depth}\\
After the initial state preparation, we ramp up the potential depth in the directions perpendicular to the 1D chains from $s_{\mathrm{L}} =\left[(30,\ 0),\ 30,\ 30\right]$ to $\left[(30,\ 0),\ 40,\ 40\right]$ in 1~ms, and then perform a sudden quench to $s_{\mathrm{L}} =\left[(s_{\text{short}}^{(x)},\ 0),\ 40,\ 40\right]$ in 0.1~ms.
Here, the tunneling time along the direction perpendicular to the 1D chains is 368~ms and the effective tunneling time between chains, derived by dividing that tunneling time by four (the number of adjacent chains), is $\hbar/(4J_{\perp})=92$~ms.
In the measurement of dependency of quench dynamics on $U/J$, the sweep time for quench is fixed to the above value independent of the potential depth along the 1D chains after the quench.
Note that we numerically confirm that even when the change in the potential depth is maximal ($U/J=12$ case), the occupation probability of the ground state of the optical lattice potential after the quench exceeds 0.99 at least for singly occupied sites.\\

\noindent
{\bf Atom number imbalance measurement}\\
After we freeze the atom distribution by rapidly ramping up the potential depth along the 1D chains in 0.1~ms to $s_{\text{short}}^{(x)}=30$, and then we ramp up the long lattice to $s_{\text{long}}^{(x)}=40$ in 1~ms, and then completely ramp down the short lattice in 1~ms.
Here, we map the particle position, i.e., the odd and even sites, to the 1st and the 3rd band of the potential of the long lattice, respectively \cite{folling2007direct} (see Fig.~\ref{Fig.1}C).
Then, the band-mapping is performed in 0.6~ms, and derive the imbalance $\mathcal{I}$ from the obtained band-mapping image.
Note that in our analysis, we count the atom numbers in the "1st Brillouin zone (BZ) $+$ half of the 2nd BZ on the closer side to the 1st BZ" and "3rd BZ $+$ half of the 2nd BZ on the closer side to the 3rd BZ" as $N_{\text{odd}}$ and $N_{\text{even}}$, respectively, taking into account the possible imperfection of our band-mapping process, although the 2nd BZ population is small.
We also note that we observe a slight deviation of imbalance from zero ($\mathcal{I}_{\text{offset}}=0.082(3)$) when we start the quench dynamics from $\ket{\cdots 1111 \cdots}$ state as a reference state which should show zero imbalance.
This might be due to the imperfection of the particle position mapping or the band-mapping.
In our analysis, we subtract this offset from the obtained imbalance $\mathcal{I}$.
Imbalance of doublons, $\mathcal{I}_\mathrm{D}$, is calculated by Eq.~\eqref{eq:I_D_def} with calibrated values of $\mathcal{I}_{\text{\ w/ PA}}$ and $\mathcal{I}_{\text{\ w/o PA}}$.

\section{Supplementary Materials}
\noindent
Supplementary material for this article is available at [URL will be inserted by publisher]\\
section~S1. Preparation of $\hat{\rho}_{\text{CDW(d)}}(0)$ state\\
section~S2. Quench dynamics starting from $\hat{\rho}_{\text{CDW(s)}}(0)$ state\\
section~S3. Bose-Hubbard parameters\\
section~S4. Scheme for removal of singlons\\
section~S5. How to prepare the initial states for the numerical calculations with matrix product states\\
section~S6. Examining effects of the density-induced tunnelings\\
fig.~S1. Schematic illustration of the preparation procedure for $\hat{\rho}_{\text{CDW(d)}}(0)$ state and $\hat{\rho}_{\text{CDW(s)}}(0)$ state in 1D chains.\\
fig.~S2. Estimation of the singlon fraction in the initial $\hat{\rho}_{\text{CDW(d,s)}}(0)$ state.\\
fig.~S3. Systematic measurement of the quench dynamics from $\hat{\rho}_{\text{CDW(s)}}(0)$ state.\\
fig.~S4. Spatial distribution of the atom number per chain $\tilde{N}_{i_y,i_z}$, where the initial state is $\rho_{\rm CDW(d,s)}(t=0)$.\\
fig.~S5. Strength of the density-induced tunneling as a function of the short-lattice depth in the axial direction $s_{\rm short}^{(x)}$ for $s_y=s_z=25$ and $40$.\\
fig.~S6. Effect of the density-induced hopping term on the imbalance dynamics.\\
table~S1. Bose-Hubbard parameters for the systematic measurement shown in Fig.~2 in the main text.\\
table~S2. Bose-Hubbard parameters for the systematic measurement shown in Fig.~S3.\\
References (64-69)

\subsection{Acknowledgments}
We thank Shintaro Taie, Luca Asteria, and Nir Navon for helpful discussions.
We also thank Yuma Nakamura for his help for the 325~nm light.
The MPS calculations in this work are performed with the ITensor library~\cite{fishman2022}.

\subsection{Funding}
This work was supported by the Grant-in-Aid for Scientific Research of JSPS [No.\ JP17H06138 (Y.Takahashi), No.\ JP18H05405 (Y.Takahashi), No.\ JP18H05228 (I.D., Y.Takahashi), No.\ JP21H01014 (I.D., Y.Takasu), No.\ JP22H05268 (M.K.), No.\ JP24KJ1332 (K.H.)], JST CREST [Nos.\ JPMJCR1673 (I.D., Y.Takahashi) and JPMJCR23I3 (Y.Takahashi)], MEXT Quantum Leap Flagship Program (MEXT Q-LEAP) [No.\ JPMXS0118069021 (I.D., Y.Takahashi)], JST Moonshot R\&D Grant [No.\ JPMJMS2268 (Y.Takahashi) and No.\ JPMJMS2269 (Y.Takahashi)], JST FOREST [No.\ JPMJFR202T (I.D.)], and JST ASPIRE [No.\ JPMJAP24C2 (M.K., I.D., Y.Takahashi)].

\subsection{Author contributions}
K.~H. and Y.~Takasu carried out experiments and the data analysis.
S.~G., H.~K., M.~K., and I.~D. carried out the theoretical calculation.
Y.~Takahashi supervised the whole project.
All the authors contributed to the writing of the manuscript.

\subsection{Competing interests}
The authors declare no competing interests.

\subsection{Data and materials availability}
All data needed to evaluate the conclusions in the paper are present in the paper and/or the Supplementary Materials.


\providecommand{\noopsort}[1]{}\providecommand{\singleletter}[1]{#1}%

\clearpage
\onecolumngrid
\begin{center}
  \textbf{\large Supplementary Materials for \\``Observation of slow relaxation due to Hilbert space fragmentation\\ in strongly interacting Bose-Hubbard chains"}  
\end{center}
\vspace{10mm}
{\bf This PDF file includes:}
\begin{itemize}
    \item section~S1. Preparation of $\hat{\rho}_{\text{CDW(d)}}(0)$ state
    \item section~S2. Quench dynamics starting from $\hat{\rho}_{\text{CDW(s)}}(0)$ state
    \item section~S3. Bose-Hubbard parameters
    \item section~S4. Scheme for removal of singlons
    \item section~S5. How to prepare the initial states for the numerical calculations with matrix product states
    \item section~S6. Examining effects of the density-induced tunnelings
    \item fig.~S1. Schematic illustration of the preparation procedure for $\hat{\rho}_{\text{CDW(d)}}(0)$ state and $\hat{\rho}_{\text{CDW(s)}}(0)$ state in 1D chains.
    \item fig.~S2. Estimation of the singlon fraction in the initial $\hat{\rho}_{\text{CDW(d,s)}}(0)$ state.
    \item fig.~S3. Systematic measurement of the quench dynamics from $\hat{\rho}_{\text{CDW(s)}}(0)$ state.
    \item fig.~S4. Spatial distribution of the atom number per chain $\tilde{N}_{i_y,i_z}$, where the initial state is $\rho_{\rm CDW(d,s)}(t=0)$.
    \item fig.~S5. Strength of the density-induced tunneling as a function of the short-lattice depth in the axial direction $s_{\rm short}^{(x)}$ for $s_y=s_z=25$ and $40$.
    \item fig.~S6. Effect of the density-induced hopping term on the imbalance dynamics.
    \item table~S1. Bose-Hubbard parameters for the systematic measurement shown in Fig.~2 in the main text.
    \item table~S2. Bose-Hubbard parameters for the systematic measurement shown in Fig.~S3.
    \item References (64-69)
\end{itemize}
\clearpage
\setcounter{section}{0}
\setcounter{figure}{0}
\setcounter{table}{0}
\setcounter{page}{1}
\renewcommand{\theequation}{S\arabic{equation}}
\renewcommand{\thefigure}{S\arabic{figure}}
\renewcommand{\thetable}{S\arabic{table}}

\section{S.1\quad Preparation of $\hat{\rho}_{\text{CDW(d)}}(0)$ state}
\begin{figure}[b]
  \includegraphics[width=8.6cm]{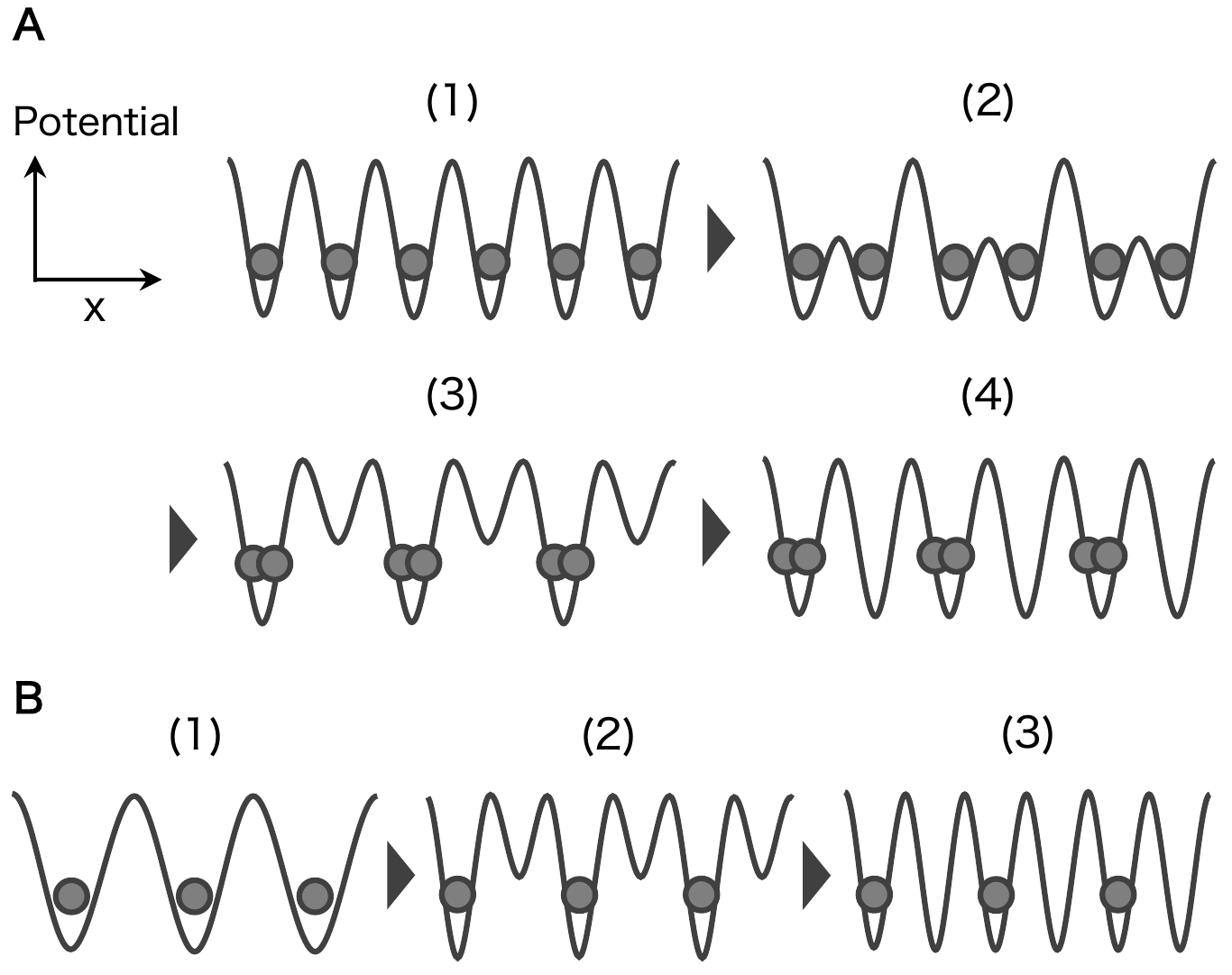}
  \caption{\label{Fig.5} {\bf Schematic illustration of the preparation procedure for (A) $\hat{\rho}_{\text{CDW(d)}}(0)$ state and (B) $\hat{\rho}_{\text{CDW(s)}}(0)$ state in 1D chains.}
  }
\end{figure}

A BEC of $^{174}$Yb is prepared by evaporative cooling in a crossed dipole trap operating at a wavelength of 532~nm, where the total atom number is about $1.3\times 10^4$.
The frequencies of the dipole trap are given by $\omega_{x+y}/(2\pi) = 27$ Hz, $\omega_{x-y}/(2\pi) = 95$ Hz, and $\omega_z/(2\pi) = 109$ Hz, respectively. Note that the axes in the $xy$ plane of the dipole trap are directed at 45 degrees to those of the optical lattice potential.
Then, we load it into a 3D cubic optical lattice with a deep potential depth of $s_{\mathrm{L}} = \left[(s_{\text{short}}^{(x)},\ s_{\text{long}}^{(x)}),\ s_{\text{short}}^{(y)},\ s_{\text{short}}^{(z)}\right]=\left[(30,\ 0),\ 30,\ 30\right]$ in 200~ms, where $s_{\text{short}}^{(x,y,z)}$ and $s_{\text{long}}^{(x)}$ are the potential depth of the short and the long lattice normalized by the recoil energy for each wavelength, respectively.
The optical lattice depth is calibrated by observing the diffraction caused by a pulsed optical lattice \cite{denschlag2002bose}.
Here, a Mott insulating state of unit filling is formed (stage (1) in Fig.~\ref{Fig.5}A). 
The optical-lattice potential at $s_{\mathrm{L}}=\left[(30,\ 0),\ 30,\ 30\right]$ also acts as a parabolic trapping potential, whose frequencies are given by $\tilde{\omega}_{x}/(2\pi) = 66$ Hz, $\tilde{\omega}_{x}/(2\pi) = 71$ Hz, and $\tilde{\omega}_z/(2\pi) = 58$ Hz.
Subsequent procedure for the $\hat{\rho}_{\text{CDW(d)}}(0)$ state preparation is as follows:

\indent
(2) Ramp up the long lattice and trap the atoms in an optical superlattice in a double-well condition along the 1D chains, where the relative phase between the short and the long lattice is $\pi/2$, and ramp down the short lattice to form the relatively shallow double-well lattice of $s_{\mathrm{L}} =\left[(10,\ 40),\ 30,\ 30\right]$ in about 20~ms.

\indent
(3) Sweep the relative phase of the short and the long lattice from $\pi/2$ to near $\pi$, or staggered condition, in 20~ms, where atoms in the right side of a double-well are transferred to the left side (stage (3) in Fig.~\ref{Fig.5}A).

\indent
(4) Ramp up the short lattice in 20~ms and then ramp down the long lattice in 1~ms to the potential depth of $s_{\mathrm{L}} =\left[(30,\ 0),\ 30,\ 30\right]$, where $\hat{\rho}_{\text{CDW(d)}}(0)$ state is to be prepared.

\begin{figure}
  \includegraphics[width=8.6cm]{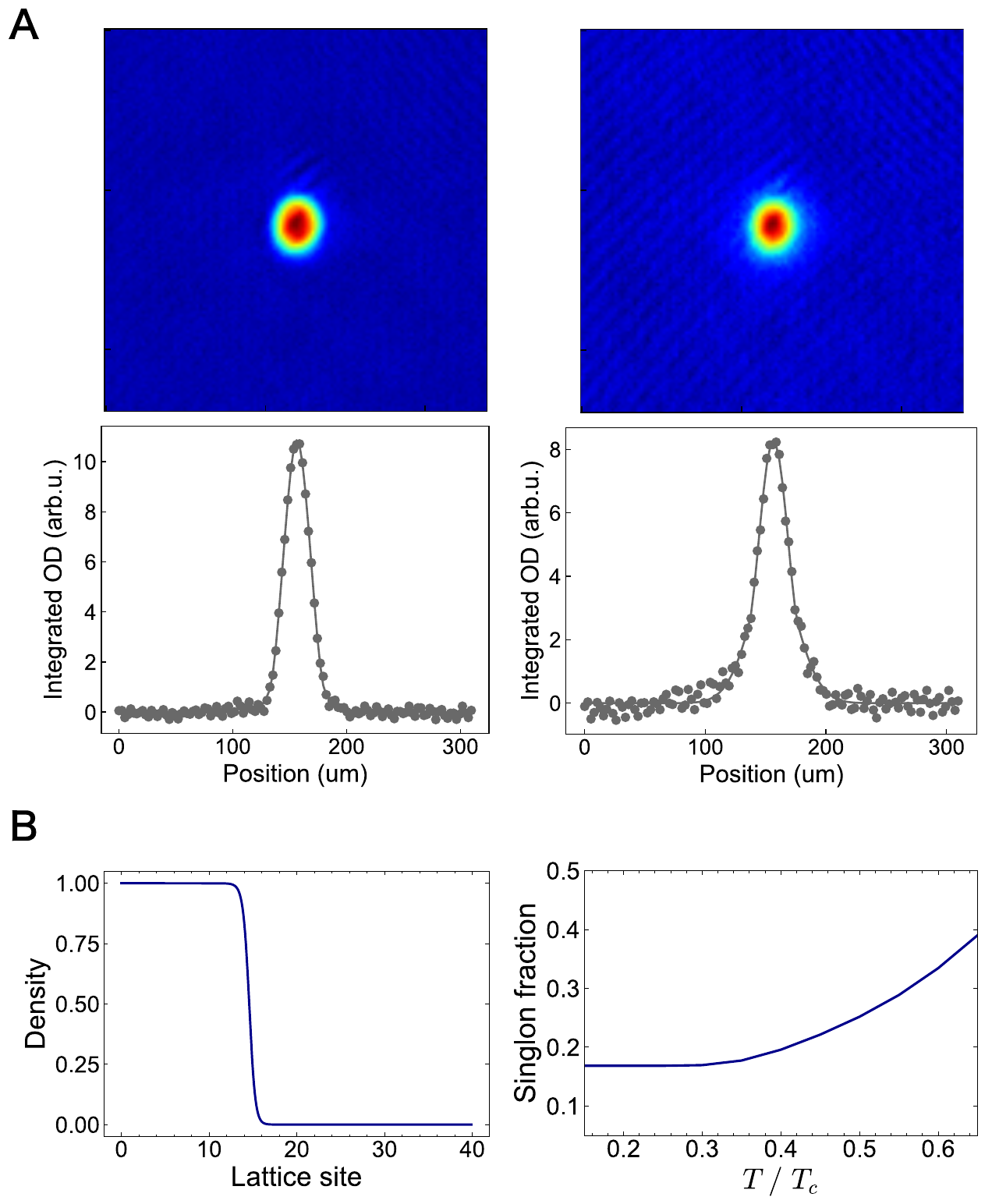}
  \caption{\label{Fig.6} {\bf Estimation of the singlon fraction in the initial $\hat{\rho}_{\text{CDW(d,s)}}(0)$ state.}
  ({\bf A})~Absorption image and its vertically integrated optical density (OD) for before (left) and after (right) `round-trip' process, taken after 14~ms time of flight.
  Both images are the average of five data.
  Solid curves in the lower left and lower right are the results of the bimodal fit.
  ({\bf B})~Calculated atom density for the temperature of 4.1~nK in the Mott insulator state, which corresponds to the entropy per particle $S/N_{\rm tot}k_{\rm B}$ of 0.23 (left), and singlon fraction as a function of the entropy per particle (right).
  }
\end{figure}

As mentioned in the main text, 25-30$\%$ of total atoms in the prepared state are singlons.
This is confirmed with the irradiation of a PA laser, where we use a PA line about 3.7~GHz below the ${^1}S_0\leftrightarrow{^3}P_1$ resonance \cite{tomita2017observation}.
Note that, we estimate the singlon fraction in the initial $\hat{\rho}_{\text{CDW(d,s)}}(0)$ state based on the atom density distribution at the Mott insulating state in $s_{\mathrm{L}} =\left[(30,\ 0),\ 30,\ 30\right]$ in the atomic limit (Fig.~\ref{Fig.6}B~(left)), where the geometric mean of the trap frequency is $2\pi \times 118$~Hz; the interaction strength is $U/h=4.74$~kHz; the total atom number $N_{\rm tot}$ is $1.3 \times 10^4$; and the temperature of the Mott insulator state is 4.1~nK.
The temperature of the Mott insulator state is obtained from $N_{\rm tot}$ and its entropy $S$, which is estimated as the half of the entropy measured after the `round-trip' process, where we first ramp up the lattice to $s_{\mathrm{L}} =\left[(30,\ 0),\ 30,\ 30\right]$ and then completely ramp down the lattice (Fig.~\ref{Fig.6}A).
We calculate the singlon fraction as $1-N_{\text{doublon}}/N_{\text{tot}}=1-2\sum_i 4\pi(2i)^2\cdot p_{2i} \cdot p_{2i+1}$, where $i$ is the lattice site index, $p_i$ is the atom density (site occupancy probability of the singlon) at a site $i$.
As a result, we obtain the singlon fraction of $20\%$ for the temperature of 4.1~nK in the Mott insulator state, which corresponds to the entropy per particle $S/N_{\rm tot}k_{\rm B}$ of 0.23, where $k_{\rm B}$ is the Boltzmann constant (Fig.~\ref{Fig.6}B~(right)).
Here, the slight deviation from the measured singlon fraction of 25-30$\%$ may be due to the nonadiabaticity in the loading process after the preparation of the Mott insulating state.

\section{S.2\quad Quench dynamics starting from $\hat{\rho}_{\text{CDW(s)}}(0)$ state}
In the measurement of the quench dynamics from the $\hat{\rho}_{\text{CDW(s)}}(0)$ state shown in Fig.~1D in the main text,
we use the following parameters in the state preparation, lattice quench, and imbalance measurement.
Recall that $\hat{\rho}_{\text{CDW(s)}}(0)$ is a mixed state created in reality when we try to prepare the ideal $\ket{\psi(0)}_{\text{CDW(s)}}=\ket{\cdots 1010 \cdots}$ state.

First, we load $^{174}$Yb BEC into a 3D cubic optical lattice with a deep potential depth of $s_{\mathrm{L}} = \left[(0,\ 40),\ 25,\ 25\right]$ in 300~ms, where the total atom number is about $4.5\times 10^3$.
Here, a Mott insulating state of unit-filling is formed (stage (1) in Fig.~\ref{Fig.5}B).
The optical-lattice potential at $s_{\mathrm{L}} = \left[(0,\ 40),\ 25,\ 25\right]$ also acts as a parabolic trapping potential, whose frequencies are given by $\tilde{\omega}_{x}/(2\pi) = 37$ Hz, $\tilde{\omega}_{x}/(2\pi) = 65$ Hz, and $\tilde{\omega}_z/(2\pi) = 53$ Hz.
Then, we ramp up the short lattice in 1~ms and trap the atoms in an optical superlattice in a staggered condition along the 1D chains, followed by ramping down of the long lattice in 1~ms, where the $\hat{\rho}_{\text{CDW(s)}}(0)$ state is prepared in the potential depth of $s_{\mathrm{L}} = \left[(25,\ 0),\ 25,\ 25\right]$ (stage (3) in Fig.~\ref{Fig.5}B).
Note that we confirm that the fraction of the doublons involved in this initial state $\ket{\psi(0)}_{\text{CDW(s)}}$ is smaller than our detection limit of about 5\% with the PA method.
For this state, we perform a sudden quench to $s_{\mathrm{L}} =\left[(s_{\text{short}}^{(x)},\ 0),\ 25,\ 25\right]$ in 0.15~ms, where the tunneling time along the direction perpendicular to the 1D chains is 38~ms.
In order to detect the imbalance after the dynamics, we freeze the atom distribution by rapidly ramping up the potential depth along the 1D chains in 0.1~ms to $s_{\text{short}}^{(x)}=25$, and then we ramp up the long lattice to $s_{\text{long}}^{(x)}=40$ in 1~ms, and finally completely ramp down the short lattice in 1~ms, followed by the band-mapping in 0.6~ms.
Note that in the imbalance measurement in the above parameters, we observe a slight deviation of imbalance from zero ($\mathcal{I}_{\text{offset}}'=0.117(3)$) when we start the quench dynamics from $\ket{\cdots 1111 \cdots}$ state, which is subtracted from the obtained imbalance.

\begin{figure}
  \includegraphics[width=6.5cm]{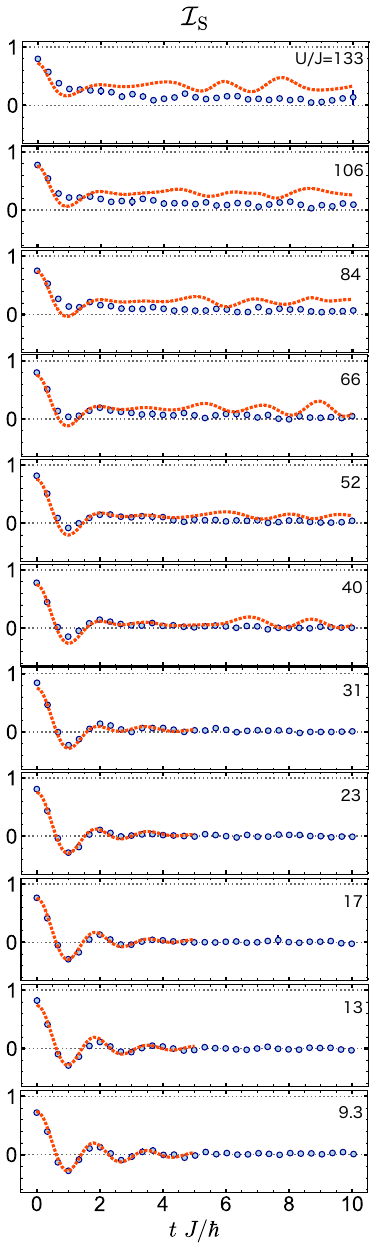}
  \caption{\label{Fig.7} {\bf Systematic measurement of the quench dynamics from $\hat{\rho}_{\text{CDW(s)}}(0)$ state.}
  The dependency of imbalance dynamics on $U/J$ is shown.
  The horizontal axis shows the holding time normalized by the tunneling time for each lattice depth along the direction of the 1D chains.
  Error bars representing the standard deviation of five independent scans are smaller than symbols.
  The dotted curves show the calculated values.
  Error bars of the theory calculations represent the statistical error and are smaller than the dots of curves.
}
\end{figure}

In Fig.~\ref{Fig.7}, we show the results of the quench dynamics from $\hat{\rho}_{\text{CDW(s)}}(0)$ state, where the dependency of imbalance dynamics on $U/J$ is systematically measured.
Here, the calculation reflects the realistic situation in the experiment, as elaborated in Sec.~S.5.
We find that the measured imbalance values are mostly reproduced by the theoretical calculations.

\section{S.3\quad Bose-Hubbard parameters}
Bose-Hubbard parameters for the systematic measurement shown in Fig.~2 in the main text 
and Fig.~\ref{Fig.7} are listed in Table~\ref{Table:Bose-Hubbard parameters_(2,0)} and Table~\ref{Table:Bose-Hubbard parameters_(1,0)}, respectively.
Here, the tunneling amplitude $J$, $U/J$, and $2J^2/U$ are obtained from the first principle calculation for an optical lattice potential, where we use the scattering length of $a_s=5.55$~nm for $^{174}$Yb atoms and three-dimensional Wannier functions to obtain the interaction strength $U$.
The harmonic potential strength $\Omega=m\omega^2d^2/2$ is obtained by 
the trap frequency along the axial direction of a 1D chain ($\omega=2\pi\times 128$~Hz in Table~\ref{Table:Bose-Hubbard parameters_(2,0)}, and $\omega=2\pi\times 109$~Hz in Table~\ref{Table:Bose-Hubbard parameters_(1,0)}), the mass $m$ of $^{174}$Yb atoms, and the lattice constant $d=266$~nm.
$\Delta\sim25\Omega$ is the maximum energy offset between neighboring sites in the parabolic trap in our system (the central 1D chains of the atoms have a length of about 25).
\renewcommand{\arraystretch}{1.5}
\begin{table}
  \centering
  \caption{{\bf Bose-Hubbard parameters for the systematic measurement shown in Fig.~2 in the main text.}\\}
  \begin{tabular}{c|c|c|c|c|c} \hline
  $s$~($E_{\rm R}$) & $J/h$~(Hz) & $U/J$ & $2J^2/U$~(Hz) & $\Omega/J$ &$\Delta/U$\\ \hline \hline
  5 & 266 & 12 & 44.3 & 0.037 & 0.077 \\
  8 & 125 & 30 & 8.33 & 0.079 & 0.066 \\ 
  10 & 77.7 & 52 & 2.99 & 0.13 & 0.063 \\
  11 & 61.9 & 67 & 1.85 & 0.16 & 0.060 \\
  12 & 49.6 & 86 & 1.15 & 0.20 & 0.058 \\
  14 & 32.4 & 138 & 0.470 & 0.31 & 0.056 \\
  15 & 26.4 & 173 & 0.305 & 0.38 & 0.055 \\ \hline
  \end{tabular}
  \label{Table:Bose-Hubbard parameters_(2,0)}
\end{table}
\renewcommand{\arraystretch}{1.}

\renewcommand{\arraystretch}{1.5}
\begin{table}
  \centering
  \caption{{\bf Bose-Hubbard parameters for the systematic measurement shown in Fig.~\ref{Fig.7}.}\\ }
  \begin{tabular}{c|c|c|c|c} \hline
  $s$~($E_{\rm R}$) & $J/h$~(Hz) & $U/J$ & $\Omega/J$ & $\Delta/U$\\ \hline \hline
  5 & 266 & 9.3 & 0.027 & 0.073 \\
  6 & 206 & 13 & 0.035 & 0.067 \\
  7 & 160 & 17 & 0.045 & 0.066 \\
  8 & 125 & 23 & 0.058 & 0.063 \\ 
  9 & 98.2 & 31 & 0.073 & 0.059 \\
  10 & 77.7 & 40 & 0.093 & 0.058 \\
  11 & 61.9 & 52 & 0.12 & 0.058 \\
  12 & 49.6 & 66 & 0.15 & 0.057 \\
  13 & 40.0 & 84 & 0.18 & 0.054 \\
  14 & 32.4 & 106 & 0.22 & 0.052 \\
  15 & 26.4 & 133 & 0.27 & 0.051 \\ \hline
  \end{tabular}
  \label{Table:Bose-Hubbard parameters_(1,0)}
\end{table}
\renewcommand{\arraystretch}{1.}

\section{S.4\quad Scheme for removal of singlons}
To investigate the effect of singlons involved in a prepared $\hat{\rho}_{\text{CDW(d,s)}}(0)$ state on the imbalance dynamics, we perform a partial removal of singlons before the quench.
Specifically, we remove only singlons from a prepared $\hat{\rho}_{\text{CDW(d,s)}}(0)$ state, which includes both doublons and singlons, by the excitation to the metastable ${^3}P_2$ state and subsequent ionization as follows (see also Fig.~4C in the main text):

\indent
(1) Irradiate the atoms in the initial state $\hat{\rho}_{\text{CDW(d,s)}}(0)$ prepared in a 3D optical lattice of $40E_{\rm R}$, where both doublons and singlons are involved, with an excitation laser resonant to the ${^1}S_0\leftrightarrow{^3}P_2\ (m_J=0)$ transition at a wavelength of 507~nm.
Here, we utilize a high-resolution laser spectroscopy with the well-separated resonance frequencies for singlons and doublons resulting from the on-site interaction shift of the resonance frequency associated with doublons, enabling us to excite only singlons to the metastable state \cite{PhysRevLett.110.173201,kato2016laser,PhysRevA.99.033609}.

(2) Irradiate the atoms with a laser beam at a wavelength of 325~nm, which induces a single-photon ionization for the metastable ${^3}P_2$ state.
Note that we observe a loss of atoms in the ${^3}P_2$ state at a rate of about 3.0~kHz for the intensity of a 325~nm laser of 23~W/$\text{cm}^2$.

In Fig.~4D in the main text,
we show the comparison of imbalance dynamics with and without partial removal of singlons.
Note that by the above scheme, we could not achieve a complete removal of singlons but only achieve a partial removal of singlons in a prepared $\hat{\rho}_{\text{CDW(d,s)}}(0)$ state, although we also try the iteration of the above removal scheme.
Although the complete removal of singlons in the initial state is not achieved in this work, as mentioned in the main text, we qualitatively confirm the effect of initially involved singlons on the quench dynamics.

\section{S.5\quad How to prepare the initial states for the numerical calculations with matrix product states}
In the experiments, a trapped atomic gas is loaded onto a deep 3D optical lattice such that the system is in a Mott insulating state where the filling factor is approximately unity, as illustrated in stage (1) of Fig.~\ref{Fig.5}A. We call this Mott insulating state as intermediate Mott state. The lattice depth for the intermediate Mott state is $s_{\mathrm{L}}=[(30,0),30,30]$ ($[(0,40),25,25]$) when we try to prepare $\ket{\psi(0)}_{\rm CDW(d)}$ ($\ket{\psi(0)}_{\rm CDW(s)}$). The fact that the initial imbalance $\mathcal{I}(t=0)\simeq 0.8$ is considerably smaller than unity means that the initial state is not perfectly the desired pure state $\ket{\psi(0)}_{\rm CDW(d)}$ or $\ket{\psi(0)}_{\rm CDW(s)}$. Moreover, when we try to prepare $\ket{\psi(0)}_{\rm CDW(d)}$, in reality a significant fraction of singlons are mixed. Since such singlons stem likely from holes in the intermediate Mott state, which are created mainly due to finite temperature effects, the presence of the singlons strongly indicates that the system is better described by a mixed state $\hat{\rho}_{\rm CDW(d,s)}(t)$ ($\hat{\rho}_{\rm CDW(s)}(t)$ when we try to prepare $\ket{\psi(0)}_{\rm CDW(s)}$).
In this appendix, we explain a phenomenological protocol, which is adopted in the numerical calculations using the MPS shown in Figs.~2 and 3 in the main text,
and Fig.~\ref{Fig.7}, for carrying out sampling of the initial pure states on the basis of the initial mixed state.

The lattice depth for the intermediate Mott state is so deep that the hopping is safely negligible. Moreover, we have confirmed that there is no detectable difference between $N_{\rm w/ \, PA}$ and $N_{\rm w/o \, PA}$ in the intermediate Mott state, meaning that we can safely neglect the occupancy of more than one atoms per site. Hence, we can approximate the density matrix of the intermediate Mott state $\hat{\rho}_{\rm MI}$ as a product state of local density matrices $\hat{\rho}_{i}$,
\begin{eqnarray}
\hat{\rho}_{\rm MI} = \bigotimes_{i}\hat{\rho}_{i},
\end{eqnarray}
where 
\begin{eqnarray}
\hat{\rho}_{i} = \sum_{n=0}^1 p_{n}^{(i)} \ket{n}\bra{n}.    
\end{eqnarray}
Here the probability $p_{n}^{(i)}$ of state $\ket{n}$ at site $i$ is given by
\begin{eqnarray}
p_{n}^{(i)}=\frac{e^{\beta\mu_i n}}{\sum_{n=0}^1 e^{\beta\mu_i n}},
\label{eq:pni}
\end{eqnarray}
where $\beta\equiv (k_{\rm B}T)^{-1}$ is the inverse temperature, and the local chemical potential $\mu_{i}$ at site $i$ is given by
\begin{eqnarray}
\mu_i &=&\mu_{\rm ctr} - \frac{1}{2}m\Biggl[ \tilde{\omega}_x^2 x_i^2 + \tilde{\omega}_y^2 y_i^2 + (\tilde{\omega}_z^2+\omega_z^2) z_i^2 
\Biggr.
\nonumber
\\
&&\left.+ \omega_{x+y}^2\left(\frac{x_i+y_i}{\sqrt{2}}\right)^2 + \omega_{x-y}^2\left(\frac{x_i-y_i}{\sqrt{2}}\right)^2\right].
\end{eqnarray}
The position of site $i$ is denoted by ${\boldsymbol r}_i=(x_i,y_i,z_i)$ and $\mu_{\rm ctr}$ means the local chemical potential at the trap center.

In order to determine the values of $\beta$ and $\mu_{\rm ctr}$, we use the two experimental observables, namely the total atom number $N_{\rm tot}$ and the number of atoms forming doublons $N_{\rm doublon}$. Since $p_{1}^{(i)}$ can be interpreted as the average density at site $i$, one can relate it to $N_{\rm tot}$ through the equation,
\begin{eqnarray}
N_{\rm tot}=\sum_{i}p_1^{(i)}.
\label{eq:Ntot}
\end{eqnarray}
Note that $p_1^{(i_x,i_y,i_z)}$ here is equivalent to $p_{i_x}$ in Sec.~S.1. 
On the other hand, $N_{\rm doublon}$ can be counted by the relation,
\begin{eqnarray}
\!\!\!\!\frac{N_{\rm doublon}}{2}=\sum_{i_x\in {\rm odd}}\sum_{i_y}\sum_{i_z}p_1^{(i_x,i_y,i_z)}p_1^{(i_x+1,i_y,i_z)},
\label{eq:Ndoublon}
\end{eqnarray}
under the following assumption. Specifically, we regard two sites of $(i_x,i_y,i_z)$ and $(i_x+1,i_y,i_z)$ as a pair, where $i_x$ is an odd integer. We assume that if both sites in a pair are singly occupied in the intermediate Mott state, a doublon is formed at the odd site after the manipulation of the secondary lattice illustrated in Fig.~\ref{Fig.5}A. Under this assumption, $p_1^{(i_x,i_y,i_z)}p_1^{(i_x+1,i_y,i_z)}$ means the probability that a doublon is formed in the pair such that the right-hand side of Eq.~(\ref{eq:Ndoublon}) counts the number of pairs accommodating doublons.  One can determine $\beta$ and $\mu_{\rm ctr}$ by substituting experimentally measured values of $N_{\rm tot}$ and $N_{\rm doublon}$ into Eqs.~(\ref{eq:Ntot}) and (\ref{eq:Ndoublon}), and solving them simultaneously. For instance, when $N_{\rm tot}=1.3\times 10^4$, $N_{\rm doublon}/N_{\rm tot}=0.775$, and $s_{\mathrm{L}}=[(30,0),30,30]$, we obtain $\beta E_{\rm R} = 20.1$ and $\mu_{\rm ctr}/E_{\rm R} = 0.326$.

With the obtained values of $\beta$ and $\mu_{\rm ctr}$, we use the probability $p_n^{(i)}$ of Eq.~(\ref{eq:pni}) to generate a spatial configuration of the atoms in the initial state of each 1D chain in the following manner. To be specific, we focus on the chain whose position in the $yz$ plane is given by $i_y$ and $i_z$. We anticipate the case that we try to prepare $\ket{\psi(0)}_{\rm CDW(d)}$ state (but fail to prepare the perfect one). First, we use random numbers to choose whether or not an atom occupies site $(i_x,i_y,i_z)$ on the basis of the probability $p_n^{(i)}$ for all $i_x$ in the chain, creating a product state, e.g., like
\begin{eqnarray}
\!\!\!\!\!\!\!\!\!\!\!\!\ket{\cdots,00,00,10,01,11,10,11,11,11,00,01,00,00,\cdots}\!\!.
\label{eq:hardcore}
\end{eqnarray}
Here we put commas in Eq.~(\ref{eq:hardcore}) in order to emphasize the pairs formed by the sites $(i_x,i_y,i_z)$ and $(i_x+1,i_y,i_z)$.
Second, we convert $\ket{11}$ state of the pairs into $\ket{20}$, immitating the state resulting from the procedure from (1) to (4) illustrated in Fig.~\ref{Fig.5}. In the case of Eq.~(\ref{eq:hardcore}), this conversion leads to
\begin{eqnarray}
\!\!\!\!\!\!\!\!\!\!\!\!\ket{\cdots,00,00,10,01,20,10,20,20,20,00,01,00,00,\cdots}\!\!.
\label{eq:doublon}
\end{eqnarray}
Third, we convert the singly occupied states ($\ket{10}$ or $\ket{01}$)  of the pairs into $\ket{01}$ with the probability $\frac{N_{\rm even}^{\rm S}}{N^{\rm S}}$ or $\ket{10}$ otherwise, where $N^{\rm S}$ is the number of atoms forming singlons and $N_{\rm even}^{\rm S}$ is that at site $i_x\in {\rm even}$. The probability $\frac{N_{\rm even}^{\rm S}}{N^{\rm S}}$ can be determined from the relation to the imbalance in the singlon sector,
\begin{eqnarray}
{\mathcal I}_{\rm S} = 1 - 2\frac{N_{\rm even}^{\rm S}}{N^{\rm S}}
\end{eqnarray}
at $t=0$. 
For instance, when ${\mathcal I}_{\rm S}= 0.4$ as is the case shown in Fig.~1 in the main text,
$\frac{N_{\rm even}^{\rm S}}{N^{\rm S}}=0.3$. 

\begin{figure}
  \includegraphics[width=8cm]{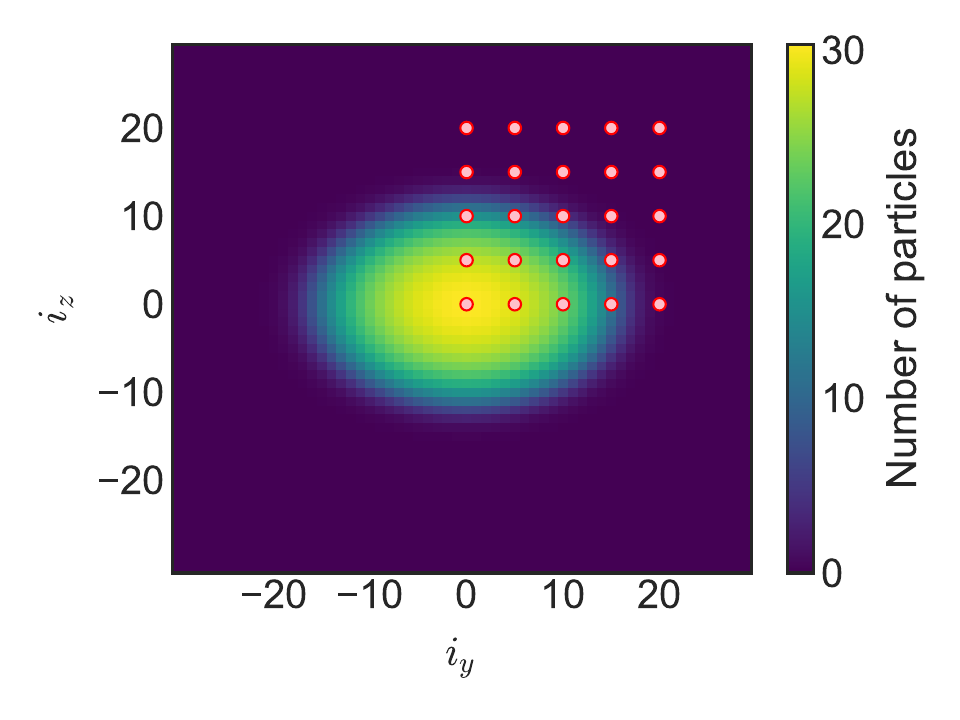}
  \caption{\label{fig:S4} {\bf Spatial distribution of the atom number per chain $\tilde{N}_{i_y,i_z}$, where the initial state is $\rho_{\rm CDW(d,s)}(t=0)$.} The open circles represent the locations of the chains $(i_y,i_z)=(5m_y,5m_z)$, where $m_y$ and $m_z$ run from 0 to $4$.
}
\end{figure}
We next consider the case that we try to prepare $\ket{\psi(0)}_{\rm CDW(s)}$ state. In this case, since $N_{\rm doublon}=0$, one can not use Eq.~(\ref{eq:Ndoublon}) for determining $\beta$ and $\mu_{\rm ctr}$ in the probability function of Eq.~(\ref{eq:pni}). Instead, by speculating the value of $\beta$ from the case of preparing $\ket{\psi(0)}_{\rm CDW(d)}$, where $\beta E_{\rm R}\simeq 20$, we set the inverse temperature to be $\beta E_{\rm R}= 20.0$. Then, solving Eq.~(\ref{eq:Ntot}) with $N=4.5\times10^3$, we obtain $\mu_{\rm ctr}/E_{\rm R}=0.198$. With these values of $\beta$ and $\mu_{\rm ctr}$, we use the probability $p_n^{(i)}$ of Eq.~(\ref{eq:pni}) to generate a spatial configuration of the atoms in the initial state of each 1D chain in a  manner similar to the case of preparing $\ket{\psi(0)}_{\rm CDW(d)}$. First, we use random numbers to choose whether or not an atom occupies site $(i_x',i_y,i_z)$ of the 3D optical lattice with $[(0,40),25,25]$ on the basis of the probability $p_n^{(i_x',i_y,i_z)}$ for all $i_x'$ in the chain, creating a product state, e.g., like
\begin{eqnarray}
\ket{\cdots,0,1,0,1,1,1,0,1,1,0,\cdots}.
\label{eq:longlattice}
\end{eqnarray}
Here, $i_x'$ denotes the site index of the $x$-direction lattice with long period. In the procedure preparing the initial state illustrated in Fig.~\ref{Fig.5}B, each site of this type is converted to two sites in the short-period lattice. Second, we convert the singly occupied state $\ket{n_{(i_x',i_y,i_z)}=1}$ into $\ket{n_{(i_x=2i_x'-1,i_y,i_z)}=0,n_{(i_x=2i_x',i_y,i_z)}=1}$ with the probability $\frac{N_{\rm even}}{N_{\rm tot}}$ or $\ket{n_{(i_x=2i_x'-1,i_y,i_z)}=1,n_{(i_x=2i_x',i_y,i_z)}=0}$ otherwise. Moreover, we convert the empty site of the long-period lattice $\ket{n_{(i_x',i_y,i_z)}=0}$ to two empty sites of the short-period lattice $\ket{n_{(i_x=2i_x'-1,i_y,i_z)}=0,n_{(i_x=2i_x',i_y,i_z)}=0}$. For instance, the state of Eq.~(\ref{eq:longlattice}) can be converted to
\begin{eqnarray}
\ket{\cdots,00,10,00,10,01,10,00,10,01,00,\cdots}.
\label{eq:shortlattice}
\end{eqnarray}

Thus, we prepare an initial state for the time evolution computed by means of the TEBD method. In order to incorporate the ‘mixedness’ of the initial state, for each 1D chain, we generate 128 samples of such an initial state and calculate the time evolution of each state independently. In order to evaluate an observable $\hat{O}$ at a certain time $t$, we take the average of $\bra{\psi(t)}\hat{O}\ket{\psi(t)}$ over the samples. 
For instance, assuming that the initial state is $\rho_{\rm CDW(d,s)}(t=0)$, we show in Fig.~\ref{fig:S4} the number of atoms in the 1D chains labeled by $(i_y,i_z)$,
\begin{eqnarray}
\tilde{N}_{i_y,i_z} = \sum_{i_x} {\rm Tr}[\hat{n}_{i_x,i_y,i_z}\rho_{\rm CDW(d,s)}(t=0)].
\end{eqnarray}
Since we neglect the interchain hoppings, $\tilde{N}_{i_y,i_z}$ at each tube is independent of time. In Fig.~\ref{fig:S4}, we see that $\tilde{N}_{i_y,i_z}$ is spatially inhomogeneous, reflecting the inhomogeneity of the trapping potentials.
Furthermore, we incorporate the spatial inhomogeneity of the atom number per chain in the $yz$ direction by taking into account several chains at $(i_y,i_z)=(5m_y,5m_z)$, where $m_y$ and $m_z$ run from 0 to $4$, and computing weighted averages of the observables with respect to the chains. These chains are marked by the open circles in Fig.~\ref{fig:S4}.
The weights are determined based on the number of equivalent chains: One for the chain at (0, 0), two for the chains at $(i_y > 0, 0)$ or $(0, i_z > 0)$, and four for the other chains.
The statistical errors of the plotted values are obtained from the standard deviation of 1024 bootstrap resampled data. Note, however, that the calculation results shown in Fig.~4 in the main text 
correspond to dynamics starting from a single pure state, and we did not use the sampling protocol for these results.

\section{S.6\quad Examining effects of the density-induced tunnelings}
\begin{figure}
  \includegraphics[width=8cm]{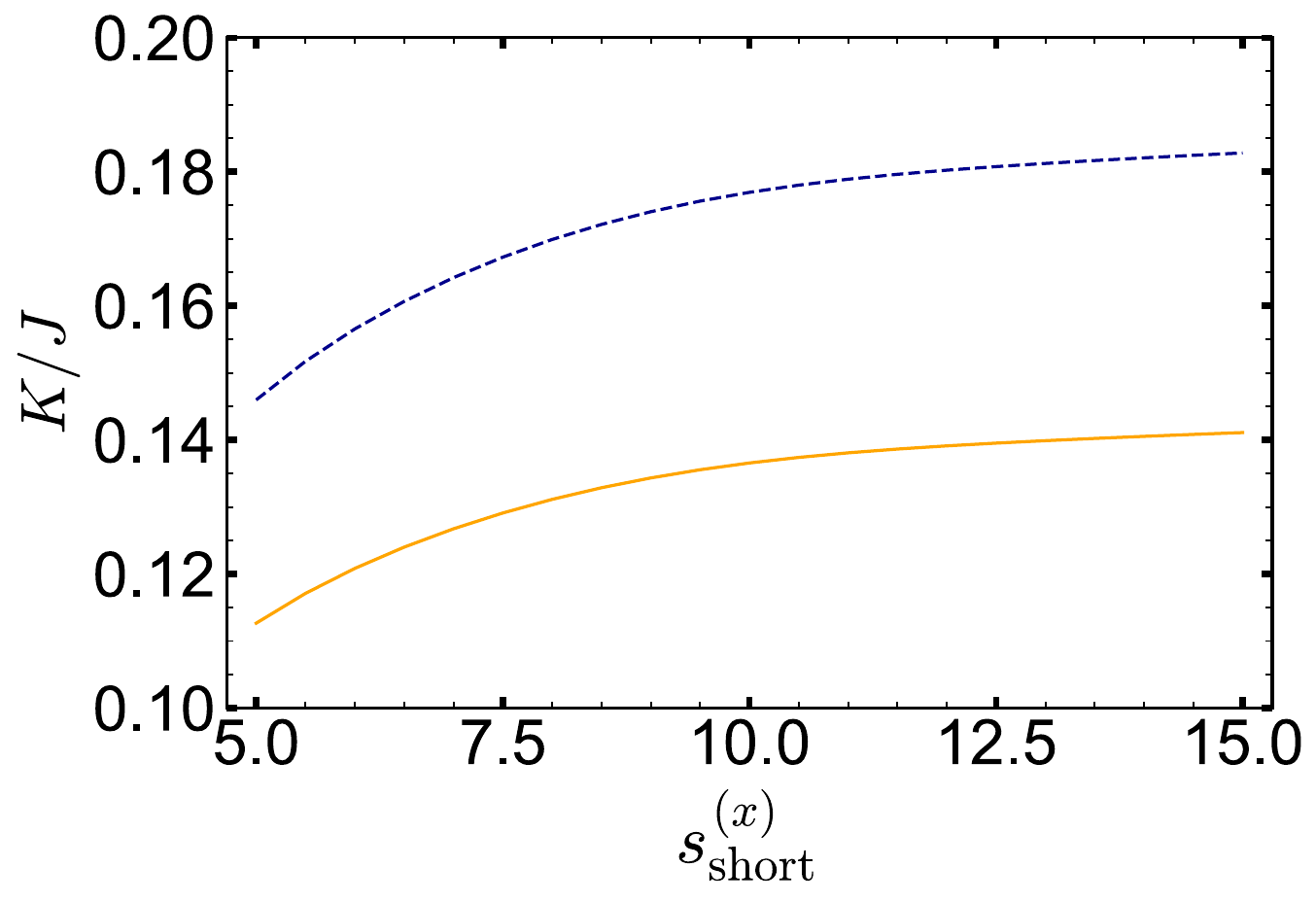}
  \caption{\label{fig:S5} {\bf Strength of the density-induced tunneling as a function of the short-lattice depth in the axial direction $s_{\rm short}^{(x)}$ for $s_y=s_z=25$ (solid curve) and $40$ (dashed curve).} Here we set the $s$-wave scattering length and the lattice spacing to be the experimental values explicitly given in Sec.~S.3, i.e., $a_s=5.55$ nm and $d=266$ nm.
}
\end{figure}

\begin{figure}
  \includegraphics[width=14.5cm]{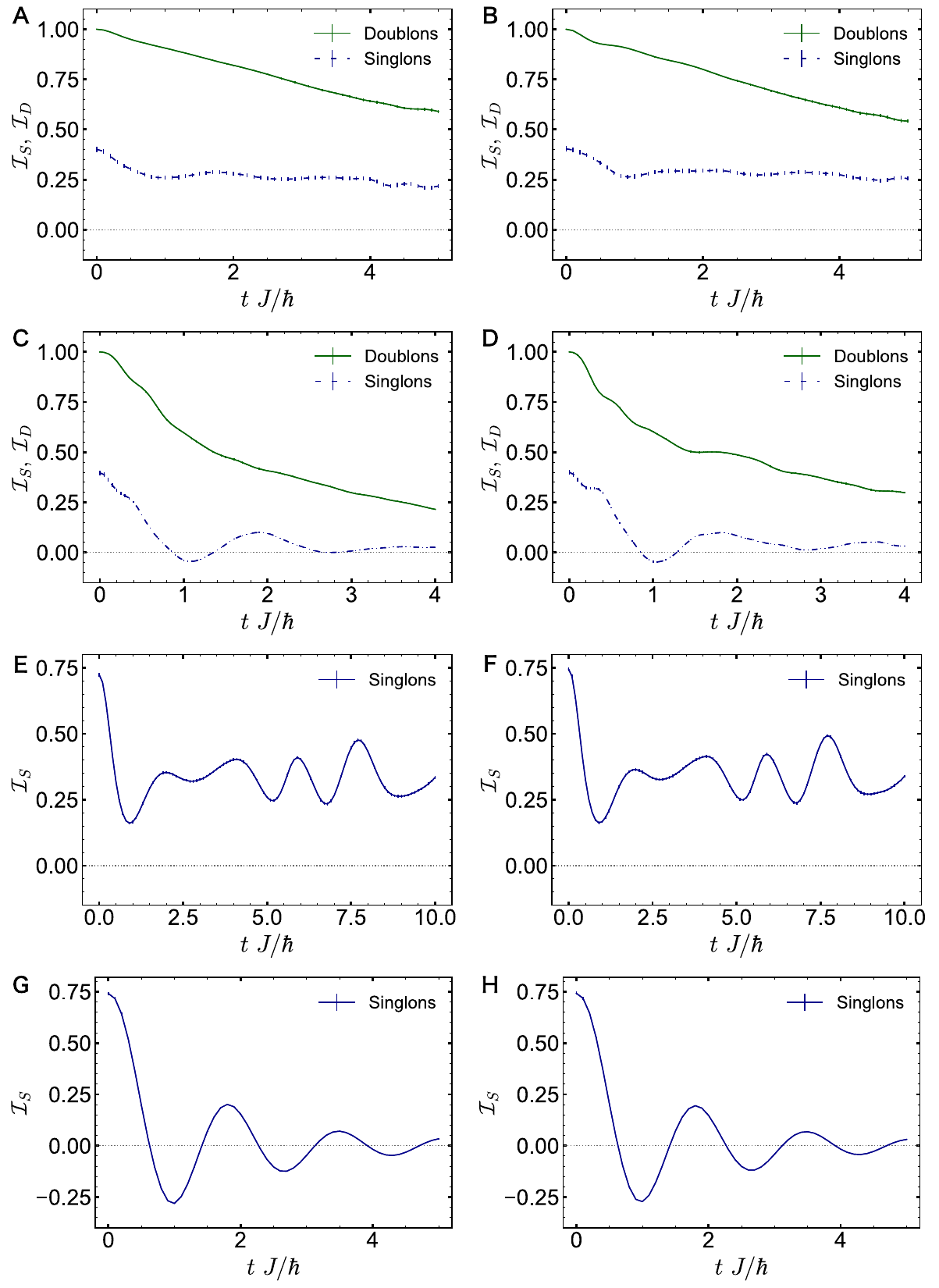}
  \caption{\label{fig:S6} {\bf Time evolution of the imbalance of doublons, and singlons in the case that the initial state is (A,B)~$\hat{\rho}_{\rm CDW(d,s)}(0)$, where $s_{\rm short}^{(x)}=15$, (C,D)~$\hat{\rho}_{\rm CDW(d,s)}(0)$, where $s_{\rm short}^{(x)}=5$, (E,F)~$\hat{\rho}_{\rm CDW(s)}(0)$, where $s_{\rm short}^{(x)}=15$, and (G,H)~$\hat{\rho}_{\rm CDW(s)}(0)$, where $s_{\rm short}^{(x)}=5$.} Left panels~(A,C,E,G): the density-induced hopping term is neglected ($K=0$), corresponding to (A,C)~the dotted lines in the top panels of Figs.~2A and 2B in the main text, and to (E,G)~the dotted lines in the top panel of Fig.~S3. Right panels~(B,D,F,H): the density-dependent hopping term is taken into account, where (B)~$K/J=0.18$, (D)~$K/J=0.15$, (F)~$K/J=0.14$, and (H)~$K/J=0.11$.
  }
\end{figure}

The numerical analyses in the main text are based on the 1D Bose-Hubbard model of Eq.~(2) in the main text, which consists of the hopping, onsite interaction, and trap potential terms. However, it is known that when the $s$-wave scattering length and/or the strength of the confinement in the radial direction is sufficiently large, the density-induced tunneling term,  
\begin{eqnarray}
\hat{H}_{\rm DIT} = - K\sum_{\langle i, j\rangle}\hat{a}_i^\dagger\left(\hat{n}_i + \hat{n}_j \right)\hat{a}_j,
\end{eqnarray}
becomes non-negligible~\cite{Dutta_2015}. In order to examine how significantly this term affects the results shown in the main text, we first depict in Fig.~\ref{fig:S5} the value of $K$ as a function of the depth of the short lattice in the $x$ direction $s_{\rm short}^{(x)}$ for the cases $s_{\rm L}=\left[\left(s_{\rm short}^{(x)},0\right),25,25\right]$ and $\left[\left(s_{\rm short}^{(x)},0\right),40,40\right]$. The former (latter) corresponds to the situation in which the initial state of the quench dynamics is $\hat{\rho}_{\rm CDW(s)}(0)$ ($\hat{\rho}_{\rm CDW(d,s)}(0)$).
Since $K/J$ is well below unity in both cases, we expect that the effects of the density-induced tunnelings should not be so significant. For confirming this expectation, we use the TEBD method combined with the initial-state preperation scheme of Sec.~S.5 in order to calculate the time evolution of the imbalance of doublons and/or singlons in the absence and presence of the density-induced tunneling term, which is shown in Fig.~\ref{fig:S6}. There we indeed see that the effects of the density-induced tunneling are rather small in the four cases, which includes both cases of the largest ($s_{\rm short}^{(x)}=15$) and smallest ($s_{\rm short}^{(x)}=5$) lattice depth among the results shown in Fig.~2 in the main text and Fig.~S3.

%
\end{document}